\documentstyle[12pt]{article}
\newcommand{\h}{\hspace*{5 ex}}
\newcommand{\disp}{\displaystyle}
\newcommand{\bi}{\bibitem}
\newcommand{\ove}{\overline}

\newcommand{\vs}{\vspace*}

\newcommand{\drm}{{\rm d}}
\newcommand{\IIrm}{{\rm II}}
\newcommand{\pa}{\partial}
\def\sottofreccia#1{\smash{\mathop{\longrightarrow}\limits_{#1}}}

\begin{document}

\centerline{\large\bf TUNNELING TIMES AND ``SUPERLUMINAL" TUNNELING:}
\vspace*{2mm}
\centerline{\large\bf A BRIEF REVIEW.$^{(*)}$}
\footnotetext{$^{(*)}$ Work partially supported by
INFN, MURST, CNR (Italy), and by the I.N.R. (Ukrainian 
Acad. Sc., Kiev).}
 
\vspace*{1 cm}
 
\centerline{Vladislav S. OLKHOVSKY$^{(a,b)}$ and Erasmo RECAMI$^{(b,c)}$}
 
\vspace*{0.5 cm}
 
{\small
\centerline{$^{(a)}$ {\em Institute for Nuclear Researches, Ukrainian
Academy of Sciences, Kiev, USSR.}}
\centerline{$^{(b)}$ {\em Facolt\`a di Ingegneria, Universit\`{a}
statale di Bergamo, Bergamo, Italy}}
\centerline{\em and I.N.F.N., Sezione di Milano, Milan, Italy.}
\centerline{$^{(c)}$ {\em C.C.S., State University at Campinas, 
Campinas, S.P., Brazil.}}
}

\vspace*{3 cm}
 
{\bf ABSTRACT --} \  In the First Part of this paper [that was submitted for
pub. in 1991 and appeared in print in Phys. Reports 214 (1992) 339] we 
critically review and analyse the main theoretical definitions and 
calculations of the sub-barrier tunnelling and reflection {\em times}.  
Moreover, we propose a new definition of such durations, on the basis of a 
recent general formalism of ours (already tested for other types of quantum 
collisions) within conventional quantum mechanics. At last, we discuss some 
surprising results regarding the temporal evolution of the tunnelling 
processes: namely, the fact that QM predicts that tunnelling through opaque 
barriers takes place with Superluminal group-velocities. \ Aims of the Second 
Part [that appeared in print later, in J. de Physique-I 5 (1995) 1351] are: \
(i) presenting and analysing the results of various numerical calculations 
(based on our equations) on the penetration and return times $<\tau_{\, 
\rm Pen}>$, $<\tau_{\, \rm Ret}>$, during tunnelling {\em inside} 
a rectangular potential barrier, for various penetration depths $x_{\rm f}$; 
 \ (ii) putting forth and discussing suitable definitions, besides of the 
mean values, also of the {\em variances} (or dispersions) ${\rm D} \, 
{\tau_{\rm T}}$ and ${\rm D} \, {\tau_{\, \rm R}}$ for the time durations of 
transmission and reflection processes; \ (iii) mentioning, moreover, that our 
definition $<\tau_{\rm T}>$ for the average transmission time results to 
constitute an {\em improvement} of the ordinary dwell--time 
${\ove \tau}^{\rm Dw}$ formula. \ The numerical evaluations {\em confirm} 
that our approach implied, and implies, the existence of Superluminal 
tunnelling (that we call ``Hartman effect"): an effect that in these days 
(due to the theoretical connections between tunnelling and evanescent--wave 
propagation) is receiving ---at Cologne, Berkeley, Florence and Vienna--- 
indirect, but quite interesting, experimental verifications. \ Eventually, 
we briefly  analyze some other definitions of tunnelling times. \ 
A more detailed review of the same topics (in Italian, however) has been
``published" electronically, as LANL Archives \# cond-mat/9802126. \ 
At last, a brief review on the experimental data, that ---in four different 
sectors of physics--- seem to indicate the existence of Superluminal motions,
appeared as an Appendix to the paper in LANL Archives \# physics/9712051
(to be published in Physica A, 1998).\hfill\break


\newpage
 
\centerline{{\large{\bf FIRST PART}}}

\

\

{\bf I-1 -- INTRODUCTION.}\\
 
\h During the last years, {\em many} attempts were
devoted to theoretically defining and calculating the time spent by a
sub-barrier--energy particle for tunnelling through a potential barrier; \
so that critical reviews were already in order. Actually, a first, valuable
review article, by Hauge and Stoveneng, did recently appear (at the end of
1989) as ref.$^{[1]}$. \ We deem it useful, however, a second review paper
since: \ (i) on one hand, it appears convenient to deepen and extend the
criticism to the existing approaches, and \ (ii) on the other hand, we can
show ---as a consequence of such a criticism--- how a definite proposal
can be put forth for the introduction of suitable, self-consistent, physically
meaningful definitions of the tunnelling times: a question that was regarded
still as an open problem in ref.$^{[1]}$.\\
\h The problem of defining the tunnelling times has a long
history, since it arose in the fortieth and fiftieth$^{[2-4]}$,
simultaneously with the problem of a general
definition in quantum mechanics of the collision durations. \ Let us explicitly
recall that the motion inside a potential barrier is a quantum phenomenon
without any classical or quasi-classical analogues, so that it lacks an easy
intuitive representation; this can explain the birth of a variety of approaches
to describe it, sometimes in contradiction with one another.\\
\h The advent of high--speed experimental devices, based on tunnelling
processes in semiconductors, did revive an interest in the whole
question;$^{[5,1]}$ \ whose relevance has always been apparent in nuclear
physics: For instance, tunnelling plays an essential role in the physics of
nuclear fission and fusion.\\
\h As already mentioned, aims of this paper are a comparative analysis of the
various, known definitions
for the tunnelling durations; and the introduction of a possible, rigorous
way for theoretically evaluating the time spent by a particle {\em inside} a
potential barrier: \ with the final presentation of some peculiarities of the
tunnelling--process time evolution. Our starting point will be the
formalism in refs.$^{[6,8]}$.\\
\h First of all, in Sect.{\bf I-2}, we critically review the main definitions
of the tunnelling times appeared up to now, to the best of our knowledge.
Then, Sect.{\bf I-3} is devoted to the application of our own
formalism$^{[6,8]}$
to tunnelling. At last, Sect.{\bf I-4} is devoted to the conclusions and to
some proposals for further developments.

\vspace*{1.5 cm}

{\bf I-2 -- ABOUT THE EXISTING DEFINITIONS OF TUNNELLING TIMES.}
 
\vspace*{0.5 cm}
 
{\bf I-2.}1 -- {\bf The simplest stationary picture of Tunnelling.}\\
 
\h Let us confine ourselves to the ordinary case of particles moving only
along the $x$--direction, and consider
a time--independent barrier $V(x)$ in the interval ($0,a$): see Fig.I-1, in
which
---for later convenience--- a larger interval ($x_{1}, x_{2}$), containing the
barrier region, is also indicated. \ \ We assume the stationary
scattering problem to have been
solved exactly for every kinetic energy
$E = \hbar^{2} k^{2}$/$2m$, where
$k$ is the wave-number and
$m$ the particle mass.  The wave function $\psi (x;k)$ will have the general
form:
 
\begin{eqnarray}
\psi \equiv \psi_{\rm I}  =  \psi_{\rm in} + \psi_{\rm R} \; \; \;
& {\rm for} & \;\;
x < 0, \nonumber\\
\psi  \equiv  \psi_{\rm II}  \;\;\;  & {\rm for} & 0 < x < a,\\
\psi  \equiv  \psi_{\rm III} = \psi_{\rm T}  \;\;\;   & {\rm for} & \;\;
x > a; \nonumber
\end{eqnarray}
\\
 
where: \  $\psi_{\rm in} \equiv
e^{ikx}$; \  $\psi_{\rm R} \equiv A_{\rm R} \; e^{-ikx}$; \ $\psi_{\rm T}
\equiv
A_{\rm T} \; e^{ikx}$; \ the lower indices R,T staying  for ``reflected" and
``transmitted", respectively. \ In the simple case of a rectangular
barrier [$V(x) = V_{0}$], it is $\psi_{\rm II} = \alpha e^{- \kappa x}
+ \beta e^{\kappa x}$ [with $\kappa \equiv  \sqrt{2m(V_{0} - E})/\hbar$],
where
the coefficients [amplitudes] $A_{\rm R}, A_{\rm T}, \alpha$ and $\beta$ can
be analytically calculated, and are known to be
 
\begin{eqnarray}
\alpha = {\frac{2k}{D_{+}}} \; {\frac{k+i\kappa}{(k^{2}-\kappa^{2})D_{-}/D_{+}
\; + \; 2ik\kappa}}; \;\;\; \beta = \frac{2k}{D_{+}} \; \frac{-k+i\kappa}
{(k^{2}-\kappa^{2})D_{-}/D_{+} \; + \; 2ik\kappa} \; \; {\rm exp}(-2\kappa a);
\nonumber\\
 \hfill \\
A_{\rm R} = \alpha + \beta -1; \;\;\;\ A_{\rm T} = (\alpha e^{\kappa a} +
\beta e^{\kappa a}) \; e^{-ka}; \;\;\; D_{\pm} \equiv 1 \pm
{\rm exp}(-2\kappa a). \nonumber
\end{eqnarray}
 
\
 
The amplitudes $A_{\rm R}$, $A_{\rm T}$ satisfy the probability conservation
law
 
\
 
\begin{equation}
|A_{\rm R}|^{2} \;\; + \;\; |A_{\rm T}|^{2} \;\;\; = \;\;\; 1 .
\end{equation}
 
\
 
\h The flux density
 
\begin{equation}
j = {\rm Re}[{\frac{i\hbar}{2m}} \psi (x) {\frac{\partial \psi^{*} (x)}
{\partial x}}]
\end{equation}
 
\
 
does not depend on $x$.  Before the barrier it equals the difference
$(1 - |A_{\rm R}|^{2})$ between the incoming and reflected wave
flux--densities. It is less known that, inside the barrier, the fluxes
for the separate components of $\psi_{\rm II}$ (exponentially decreasing and
increasing, respectively:  $\alpha
e^{-\kappa x}$ and $\beta e^{\kappa x}$) do vanish.  Only their interference
does provide the conservation of $j$.\\
 
\newpage
 
{\bf I-2.}2 --  {\bf Construction of the wave--packet}\\
 
\h For later use in the non-stationary description of actually moving
wave--packets, let us consider a wave--packet constructed in terms of
the solutions $\psi (x;k)$ of the stationary Schroedinger equation:
namely, by integrating over $E$ from $0$ to $\infty$  with a
weight-amplitude $g(E - \ove{E})$\\
 
\begin{equation}
\Psi (x,t) = \int {\rm d}E \; g(E-\ove{E}) \; \psi (x;k) \;
{\rm exp}(-iEt/\hbar)
\end{equation}
 
\
 
where we introduced the resolution of the time-evolution operator, with the
normalization condition  $\int {\rm d}E \; |g|^{2} = 1$, quantity $\ove{E}$
being the average kinetic energy.\\
\h In the case of a Gaussian wave--packet it is convenient to pass from the
energy to the impulse representation, by recalling that ${\rm d}E =
(\hbar^{2} k/2m) {\rm d}k$; \ when the spread in $E$ of
$g(E-\ove{E})$ is much smaller than $\ove{E}$, one easily gets
 
\
 
\begin{equation}
{g(E-\ove{E}) \; {\rm d}E} \approx {G(k-\ove{k}) \; {\rm d}k} \equiv
{{\frac{\hbar^{2} \ove{k}}{2m}} \; \; g(\frac{\hbar^{2}}{2m} \; [k^{2} -
{{\ove{k}}^{2}}]) \;\, {\rm d}k} ,
\end{equation}
 
\
 
with $G \equiv C \; {\rm exp}[-b(k-\ove{k})^{2}]$. \ Of course, the
(initial) wave--packet of the incoming waves will have a Gaussian shape
also in the configuration space.\\
\h By inserting in the integral of eq.(5) the reflected ($\psi_{\rm R}$) or
transmitted  ($\psi_{\rm T}$) wave, instead of the total wave $\psi$, we
obtain the {\em final} reflected or transmitted wave--packets, respectively,
carrying a time-delay due to the interaction.  Notice that one could expect a
distortion in the wave--packet form due to the energy dependence of $A_{\rm R}$
and $A_{\rm T}$; but it has been already shown, for a wide class of
weight amplitudes, such a distortion to be negligible.$^{[7]}$  Furthermore,
we shall get rid also of the wave components with above-barrier energies by
introducing the additional transformation
 
\begin{equation}
g(E-\ove{E}) \longrightarrow g(E-\ove{E}) \; \Theta (E-V_{0}) ,
\end{equation}
 
\
 
in order to avoid distortions of the sub-barrier penetration (tunnelling).\\
\h For simplicity's sake, we shall in general address ourselves to
quasi-monochromatic packets, for which the energy spread $\Delta E$ is
so much smaller than $\ove{E}$ that it is possible to adopt the approximation
 
\begin{equation}
|g(E-\ove{E})|^{2} \simeq \delta (E-\ove{E})
\end{equation}
 
\
 
in all the final expressions for our quantities, when averaged over
$\rho {\rm d}x$ or $J {\rm d}t$; where
 
\
 
\begin{equation}
\rho \equiv |\Psi (x,t)|^{2} ; \; \; \; J \equiv
{\rm Re}[{\frac{i\hbar}{2m}} \Psi (x,t) {\frac{\partial \Psi^{*} (x,t)}
{\partial x}}]
\end{equation}
 
\
 
are the probability density [for a particle to be located in the unitary
space--interval centered at $x$] and the probability--density flux
[for a particle
to pass through position $x$ during the unitary time--interval centered
at $t$], respectively.
 
\vspace*{0.75 cm}
 
{\bf I-2.}3 --  {\bf The ordinary phase--times}\\
 
\h Following the usual procedure, introduced in$^{[2-4]}$, it is easy to
get the ordinary phase--times for quasi-monochromatic packets, in the
stationary--phase approximation
 
\begin{equation}
\tau_{\rm T}^{\rm Ph} (x_{\rm i}, x_{\rm f};\; E) = \frac{1}{v} \; (x_{\rm f}
- x_{\rm i}) + \hbar \; \frac{{\rm d}(\arg A_{\rm T})}{{\rm d}E}
\end{equation}
 
and
 
\begin{equation}
\tau_{\rm R}^{\rm Ph} (x_{\rm i}, x_{\rm i}; E) = \frac{1}{v} \;
(-2x_{\rm i}) + \hbar \; \frac{{\rm d}(\arg A_{\rm R})}{{\rm d}E}
\end{equation}
 
\
 
where $v \equiv \hbar k/m$ is the group--velocity; $x_{\rm i} \in$ region I,  
\ and $x_{\rm f} \in$ region III. \  Eqs.(10) and (11) refer to
a transmitted [from the initial position $x_{\rm i}$ to the final position
$x_{\rm f}$] particle and to a reflected [from the initial position $x_{\rm i}$
to the same position] particle, respectively; cf., {\em e.g.}, ref.$^{[1]}$.\\
\h For a rectangular barrier with height $V_{0}$, the phase--times (10)
and (11),
when linearly extrapolated$^{[1]}$ to the barrier region ($x_{\rm i} = 0;
\; x_{\rm f} = a$) would become
 
\
 
\begin{equation}
\tau_{\rm T}^{\rm Ph} (0,a;E) = \tau_{\rm R}^{\rm Ph} (0,a;E) =
\frac{m}{\hbar
k \kappa D} \; [2 \kappa a k^{2} (\kappa^{2} - k^{2}) + {k_{0}}^{4}
{\rm Sinh}(2 \kappa a)] \; ,
\end{equation}
 
\
 
which, for $\kappa a >> 1$, would simply yield $2/v \kappa$. In eqs.(12), it
is $D \equiv 4 \kappa^{2} k^{2} + {k_{0}}^{4} {\rm Sinh}^{2}(\kappa a)$; and
$k_{0} \equiv 2m{V_{0}}/\hbar$. \   In other words,$^{[7]}$
for sufficiently wide ---{\em i.e.}, opaque--- (or high) barriers, eqs.(12)
do not depend on the
barrier width $a$, and the effective tunnelling--velocity $a/\tau^{\rm Ph}$
may become arbitrarily large [Hartmann and Fletcher effect$^{[9,10]}$].\\
\h One of the main objections against extrapolations (12) is that
they do not describe the actual asymptotic behaviour of the phase--times;
since they disregard the fact that both the [magnitude of the]
initial packet mean--position, $|x_{\rm i}|$, and quantity $x_{\rm f} - a$
(where $x_{\rm f}$ is the transmitted packet mean--position) must be
{\rm large} with respect to the packet spatial extension [of the order of
$\hbar v/\Delta E$], in order to avoid ``interference" effects between
physically quite different processes ({\em i.e.}, between incident and
reflected waves).\\
\h Therefore, it is not completely correct to attribute to the
extrapolated phase--times the physical meaning of ``times spent in the
barrier region (= inside the barrier)".  Moreover, one cannot separate
in $\tau_{\rm T}^{\rm Ph}$ and $\tau_{\rm R}^{\rm Ph}$ the self-interference
delays from the time spent inside the barrier.\\
\h Before going on, let us clarify the behaviour of the phase--times
at the very {\em top} of the barrier, and check whether there is any
continuity ---there--- between the values of the sub-barrier
tunnelling time and those for the above-barrier case. \ Let
us compare eqs.(12) with the following expression for the above-barrier
transmission time:
 
\
 
\begin{equation}
\tau_{\rm T}^{\rm Ph} (0,a;E>V_{0}) =  {\frac{2m}{\hbar k q}} \;\;
{\frac{-(k^{2}-q^{2})^{2}{\rm tan}(qa) + 4qak^{2}(k^{2}+q^{2})/{\rm cos}^{2}
(qa)}{4k^{2}q^{2} + [(k^{2}+q^{2}){\rm tan}(qa)]^{2} }}
\end{equation}
 
\
 
which was obtained$^{(*)}$,
\footnotetext{$^{(*)}$ These calculations have been explicitly performed
by V.S. Sergeyev.}
by the stationary--phase method, for the case of a rectangular
barrier. In such a case, it is \ $\psi_{\rm II} = \gamma e^{iqx} +
\delta e^{-iqx}$ \ with
$q \equiv \sqrt{2m(E-V_{0})}/\hbar$, and the coefficients $\gamma$ and
$\delta$ are analytically evaluable. \ By comparing eqs.(12) and (13) one
gets
 
\
 
\
 
\[
\lim_{\kappa \rightarrow 0} \tau_{\rm T}^{\rm Ph} (0,a;E>V_{0})
= \frac{mka^{3}}
{6\hbar (1+k^{2}a^{2}/4)} \;\;\; \sottofreccia{a \rightarrow \infty}
\;\;\; \frac{2ma}{3\hbar k} \; , \]  \nolinebreak  \hfill  (12')
 
\
 
\[
\lim_{\kappa \rightarrow 0} \tau_{\rm T}^{\rm Ph} (0,a;E<V_{0}) =
\frac{mka^{3}}
{6\hbar (1+k^{2}a^{2}/4)} \;\;\; \sottofreccia{a \rightarrow \infty}
\;\;\; \frac{2ma}{3\hbar k} \; . \]  \nolinebreak  \hfill  (13')
 
\
 
\
 
In other words, we find that that the two limits (12'), (13') do
coincide with each other,
and linearly depend on $a$ for ``opaque" barriers (provided that the
condition $\kappa a \rightarrow 0$ holds). Notice that such a result does not
contradict the Hartmann and Fletcher effect, since the latter takes place
only when $\kappa a \rightarrow \infty$, while it is absent for finite
values of $\kappa a$.\\
 
\setlength{\textheight}{22 cm}

\newpage
 
{\bf I-2.}4 --  {\bf The dwell time}\\
 
\h The total scattering time duration has been defined in$^{[11]}$ as the
probability for the particle to be localized in the interval
between the initial (maybe, source) position and the final (maybe, detector)
position, divided by the incident particle flux density; that is to say, as
the time spent by a particle while travelling inside such space--interval:
the so--called {\em dwell time}.  In the chosen case of particles moving
only along $x$, the dwell
time is therefore defined as:$^{[12]}$
 
\begin{equation}
\tau^{\rm Dw} (x_{\rm i}, x_{\rm f}; k) = \frac{1}{v} \; \int_{x_{\rm i}}
^{x_{\rm f}} {\rm d}x \; |\psi(x;k)|^{2} .
\end{equation}
 
\
 
\h For a rectangular barrier, the (dwell) time spent inside the barrier
becomes:$^{[12]}$
 
\
 
\begin{equation}
\tau^{\rm Dw} (0, a; k) = \frac{mk}{\hbar
\kappa D} \; [2 \kappa a  (\kappa^{2} - k^{2}) + \frac{2mV_{0}}{\hbar^{2}}
{\rm Sinh}(2 \kappa a)]
\end{equation}
 
\
 
which, for $\kappa a >> 1$, would give $\hbar k/\kappa V_{0}$. The results (12)
and (15) are in sharp contrast with each other with regard to the
$k$-dependence. Let us comment on this point.\\
\h The stationary definition (14) for the dwell time, according to us,
is not self-consistent from its very beginning, and appears to be in
contradiction with its physical meaning.  In fact, the time variable is
firstly discarded (in passing from the time--dependent to the stationary
Schroedinger equation), and later on it is re-introduced in an artificial,
{\em ad hoc} way: namely, through the introduction of a
localization--probability expressed in terms of time--independent
wave functions, instead of actually moving wave--packets.\\
\h Moreover, even
the modified ``dwell time approaches" with time--dependent wave
functions$^{[13-15]}$, in which
 
\
 
\begin{equation}
{\ove\tau}^{\rm Dw}(x_{\rm i},x_{\rm f};k) = \frac{\int_{-\infty}^{\infty}
{\rm d}t \int_{x_{\rm i}}^{x_{\rm f}} {\rm d}x \; \rho (x,t)}
{\int_{-\infty}^{\infty} {\rm d}t \; J_{\rm in}(x_{\rm i},t)} \;\; ,
\end{equation}
 
\
 
do still contain {\em formal} time--averages, that are {\em not} actual averages
over the physical time ({\em i.e.}, the time $t(x)$ at which the considered
particle passes through the position $x$). In fact 
those averages should be obtained
---at least when the direction of flux $J$ is time--independent--- by
integrating $J{\rm d}t$, and not $\rho {\rm d}t$.$^{[6,16]}$  \  In eq.(16)
quantity $J_{\rm in}$ is defined as in eq.(9), just replacing $\psi (x;k)$
of eq.(5) by $\psi_{\rm in}(x;k) \equiv e^{ikx}$. \  At last, the
``dwell--time approaches" are unable$^{[6]}$ to define \linebreak
\nopagebreak and study the
time {\em distributions} for any kind of collision process.\\

\setlength{\textheight}{21 cm}

\newpage

{\bf I-2.}5 --  {\bf The local Larmor times}\\
 
\h In$^{[17]}$ a {\em gedanken experimente} was proposed for measuring the
scattering
duration as the ratio $\theta/\omega$, where $\theta$ is the angle by which the
magnetic moment ${\bf \mu}$ of the considered particle is rotated due to a small
homogeneous magnetic
field ${\bf B} = B_{0}${\bf {\^{z}}} (directed along $z$) supposedly present
in the scattering region,
and $\omega \equiv 2\mu B_{0}/\hbar$ is the Larmor precession frequency.  For
a magnetic field existing in the interval ($x_{\rm i},x_{\rm f}$), and for an
incident particle (moving along $x$ and) with spin $\frac{1}{2}$ polarized
along the $x$-direction
(see Fig.I-2), $\theta$ results to be proportional to the average spin component
$<s_{y}>$: namely, $\theta = -2{{<s_{y}>}_{\rm T}}/\hbar$, \ or \ $\theta =
-2{{<s_{y}>}_{\rm R}}/\hbar$, for the transmitted or reflected waves,
respectively.  In this case, the Larmor times \  $\tau_{y {\rm T}}^{\rm La}
(x_{\rm i},x_{\rm f};k)$ \ and \  $\tau_{y {\rm R}}^{\rm La}
(x_{\rm i},x_{\rm f};k)$ \  become equal to the Phase Times, {\em plus}
---however--- terms which do oscillate as $kx_{\rm i}$ and $kx_{\rm f}$
vary.$^{[1]}$ \  In the particular case of a rectangular barrier one gets
 
\
 
\begin{equation}
\tau_{y {\rm T}}^{\rm La}(0,a;k) = \, \tau_{y {\rm R}}^{\rm La}(0,a;k) =
\tau^{\rm Dw}(0,a;k) .
\end{equation}
 
\
 
\h In Baz's approach, as it was shown in refs.$^{[6]}$, the expressions for
the collision duration [{\em e.g.}, eqs.(17)] are artificially distorted
by the sharp boundaries attributed to the magnetic--field region; in other
words, are influenced by the mathematical, rather than physical, assumptions.
Actually, the oscillating terms do depend on the kind of boundary
[for instance, smoothed] that one adopts. Moreover, both those
oscillating terms vanish when $x_{\rm i} \rightarrow -\infty$ and
$x_{\rm f} \rightarrow \infty$, once one does average over the incident
particle energy--spread; so that the final expressions do coincide with
the Phase Times.\\
\h It it also known$^{[1]}$ that the mathematical behaviour assumed for the
magnetic--field boundary does not influence only the
spin components along $y$. In fact (see Fig.I-2), the incident particle has
finite probabilities of being spin-up or spin-down along the field--direction
$z$.  As pointed out in ref.$^{[12]}$, the spin-up components will be
preferentially {\em transmitted} (except when ${\rm d}|A_{\rm T}|^{2}/
{\rm d}E < 0$): so that one gets the noticeable result that
${<s_{z}>}_{\rm T} \;\;\; >> \;\;\; {<s_{y}>}_{\rm T}$. \ On the basis of
what precedes, B\"{u}ttiker in ref.$^{[12]}$ introduced the new Larmor
times: \ $\tau_{z {\rm T}}^{\rm La}(x_{\rm i},x_{\rm f};k)$ \ and \
$\tau_{z {\rm R}}^{\rm La}(x_{\rm i},x_{\rm f};k)$, \ both defined analogously
to quantities $\tau_{y {\rm T}}^{\rm La}$ and $\tau_{y {\rm R}}^{\rm La}$,
respectively; as well as the {\em hybrid} Larmor times, defined as follows:
 
\
 
\begin{equation}
(\tau_{\rm T,R}^{\rm La})^{2} = (\tau_{y {\rm T,R}}^{\rm La})^{2} +
(\tau_{z {\rm T,R}}^{\rm La})^{2} .
\end{equation}
 
\
 
\h However, the introduction of so many time durations for a single collision
({\em e.g.}, transmission and  reflection processes) seems to us physically
unjustified.\\
\h Let us notice that, for an {\em opaque} rectangular barrier, we obtain
 
\
 
\begin{equation}
\tau_{z {\rm T}}^{\rm La}(0,a;k) \simeq \frac{ma}{\hbar \kappa} \;\; ,
\end{equation}
 
\
 
which results to be different from both the extrapolated Phase Times (12')
and the Dwell Times (15).
 
\vspace*{0.75 cm}
 
{\bf I-2.}6 --  {\bf A complex time approach}\\
 
\h As it is known, a formal generalization of the classical time spent by a
particle inside
the barrier can lead, for $E < V_{0}$ (when the actual presence of the
particle, there, is forbidden by classical mechanics), to the introduction of a
complex time.  More generally, an analogous extension of classical time {\em
to the quantum domain} has been recently proposed in$^{[18]}$ (see also,
{\em e.g.}, refs.$^{[19-21]}$, and refs. therein).  For one--dimensional
motion, following ref.$^{[18]}$,
one of the natural ``quantum generalizations" of the classical expression
 
\
 
\begin{equation}
\tau [P(t)] = \int_{t_{\rm i}}^{t_{\rm f}} {\rm d}t \int_{V}
{\rm d}x \;\; \delta [x-P(t)]
\end{equation}
 
\
 
for the time spent inside a region $V$ ---where $P(t)$ is the
classical path going from $x_{\rm i}(t_{\rm i})$ to
$x_{\rm f}(t_{\rm f})$--- is the path--integral average
 
\
 
\begin{equation}
\tau^{\rm Qu}(x_{\rm i},t_{\rm i};x_{\rm f},t_{\rm f}) \, =
\; <\tau [P(\;\;)]>_{\rm paths} \;\; ,
\end{equation}
 
\
 
in which $P(\;\;)$ is any arbitrary path between the given end--points.\\
\h For the process relative to Fig.I-1 and to eq.(1), one has:$^{[5,19]}$
 
\
 
\begin{equation}
\tau_{\rm T}^{\rm Qu} = i\hbar \; \int_{V} {\rm d}x \; \frac
{\delta \, {\rm log}
A_{\rm T}}{\delta \Omega (x)} \; ; \;\;\;\;\;\; \tau_{\rm R}^{\rm Qu} =
i\hbar \; \int_{V} {\rm d}x \; \frac{\delta \, {\rm log}
A_{\rm R}}{\delta P(x)} \;\; ,
\end{equation}
 
\
 
where $V$ is nothing but the interval ($x_{\rm i}, x_{\rm f}$) [or, in
particular,
($0, a$)]; and $\delta / \delta P(x)$ is a functional derivative. \ In general,
quantities $\tau_{\rm T,R}^{\rm Qu}$ are complex; and are connected with the
Larmor Times by the relations
 
\
 
\begin{equation}
{\rm Re} \; \tau_{\rm T,R}^{\rm Qu} = \tau_{y {\rm T,R}}^{\rm La} \; ;\;\;\;
\;\;\;\;\;\; {\rm Im} \; \tau_{\rm T,R}^{\rm Qu} =
- \tau_{z {\rm T,R}}^{\rm La} \;\; .
\end{equation}
 
\
 
\h Of course, complex time is a useful theoretical tool; even if the ordinary
tunnelling--times should be real.  The physical meaning of the imaginary
part is still controversial.$^{[22]}$
 
\vspace*{0.75 cm}
 
{\bf I-2.}7 --  {\bf The B\"{u}ttiker--Landauer time}\\
 
\h In refs.$^{[23]}$ the tunnelling times were studied via a new kind of
``gedanken experimente", namely by supposing the barrier to possess,
besides the ordinary (time--independent) part, an additional part
oscillating in time:
 
\
 
\begin{equation}
V(t) = V_{0} + V_{1} \; {\rm cos} \, \omega t \;\; .
\end{equation}
 
\
 
Since the potential $V$ varies with time, the incident particles ---if endowed
with electric charge (or magnetic moment)--- can absorb
or emit "modulation quanta" $\hbar \omega$ during the tunnelling, which leads to
the appearance of {\em sidebands} with energies $E + n \hbar \omega$; $\;$
[$n = \pm 1, \pm 2, ...$]. \  In the first--order approximation in $V_{1}$, it
is enough to consider only the neighboring sidebands with energies
$E \pm \hbar \nu$.  B\"{u}ttiker and Landauer did obtain the following
expressions for the relative sideband intensities

\begin{eqnarray}
{I_{\rm T}^{(\pm 1)}}(\omega) & = & \mid {\frac{A_{\rm T}^{(\pm 1)}(\omega)}
{A_{\rm T}^{(0)}}}\mid^{2} \; \simeq \; [\frac{V_{1}}{2\hbar \omega} \;
\exp (\pm \omega \tau^{\rm BL}_{\rm T}) - 1]^{2} \; , \nonumber \\
\hfill \\
{I_{\rm R}^{(\pm 1)}}(\omega) & = & \mid {\frac{A_{\rm R}^{(\pm 1)}(\omega)}
{A_{\rm R}^{(0)}}}\mid^{2} \; \simeq \; (\frac{V_{1} \tau^{\rm BL}_{\rm R}}
{2 \hbar})^{2} (1 \pm \omega \tau_{\kappa}) \; , \nonumber
\end{eqnarray}
 
\
 
where $A_{\rm T}^{(\pm 1)}$ and $A_{\rm T}^{(0)}$ are the perturbed (sideband)
and unperturbed transmission--amplitudes, respectively; and similarly for
the reflection--amplitudes $A_{\rm R}^{(\pm 1)}$ and $A_{\rm R}^{(0)}$. \
In eqs.(25) the last equalities ($\simeq$) hold only for the case of
{\em opaque} [rectangular] barriers and not too high frequencies: {\em i.e.},
for $\hbar \omega$ small with respect to both $E$ and $V_{0} - E$. \
Moreover, $\; \tau^{\rm BL}_{\rm T} \equiv ma/\hbar \kappa$; \ \
$\tau^{\rm BL}_{\rm R} \equiv 2mk/[\hbar \kappa (\kappa^{2} +
k^{2})]$; \ and \ $\tau_{\kappa} \equiv m/\hbar \kappa^{2}$. \ One can see
that $\tau^{\rm BL}_{\rm T}$ is identical to the Larmor time
$\tau^{\rm La}_{z {\rm T}}$ as given in eq.(19).\\
\h In our opinion, it is not worthwhile to report about the discussions
originated by B\"{u}ttiker--Landauer's approach, since they seem to us
as being too technical and insufficiently justified; let us only quote, here,
the refs.$^{[24-27]}$ \ \ However, two results should be mentioned.\\
\h First, Hauge and
Stovneng$^{[1]}$ did find a
simple connection between, on one side, the $\omega \rightarrow 0$
limits of $A_{\rm T}^{(\pm 1)}(\omega)/A_{\rm T}^{(0)}$ and
$A_{\rm R}^{(\pm 1)}(\omega)/A_{\rm R}^{(0)}$, and, on the other side,
the complex times of eqs.(23):
 
\
 
\begin{equation}
\frac{A_{\rm R}^{(\pm 1)}(\omega)}{A_{\rm R}^{(0)}} = -i \, \frac{V_{1}}{2\hbar}
\; \tau^{\rm Qu}_{\rm T} \; ; \;\;\;\;\;\; \frac{A_{\rm T}^{(\pm 1)}(\omega)}
{A_{\rm T}^{(0)}} = -i \, \frac{V_{1}}{2\hbar}\; \tau^{\rm Qu}_{\rm R} \; ,
\end{equation}
 
\
 
even if the physical meaning of such a connection is not yet very clear.\\
\h Second,  it is
interesting to recall that Bruinsma and Bak$^{[28]}$ (see also
ref.$^{[1]}$) proposed the characteristic frequency \ $(\tau^{\rm BL}_{\rm T})
^{-1} \equiv \hbar \kappa/ma$ \ to give information about the {\em coupling}
between
tunnelling and other accompanying channels, rather than about the intrinsic
tunnelling times.
 
\vspace*{1.5 cm}

{\bf I-3 -- ABOUT THE POSSIBILITY OF INTRODUCING CLEAR DEFINITIONS OF
$\tau_{\rm T}$ AND $\tau_{\rm R}$.}
 
\vspace*{0.5 cm}
 
{\bf I-3.}1 -- {\bf A comment on Hauge and Stovneng's conclusions.}\\
 
\h After having reviewed the main definitions and evaluations of the
tunnelling times (which we also have presented, and criticized, in
Sect.{\bf I-2}), the authors of ref.$^{[1]}$ concluded that no definite,
acceptable approach still exists to calculating such  tunnelling
durations.  As a necessary but not sufficient
condition, to be obeyed by any physically acceptable expression of the
tunnelling and reflection times $\tau_{\rm T}$ and $\tau_{\rm R}$, those
authors did propose the following relation [T = Transmitted, in this case!]
 
\
 
\begin{equation}
\tau^{\rm Dw} = |A_{\rm T}|^{2} \tau_{\rm T} \; + \; |A_{\rm R}|^{2}
\tau_{\rm R} \; ,
\end{equation}
 
\
 
which they required to be satisfied by the durations calculated via
{\em any} method (except the dwell one, of course, which {\em ab initio}
does not separate transmission from reflection time). \ Let us observe that
the negative conclusion of ref.$^{[1]}$, which is
actually the main conclusion of that review, is based not only on a
criticism of all the previously existing approaches (a criticism that we
made more complete and even stronger), but also on the fact that none of
them satisfies condition (27).\\
\h However, relation (27) is unacceptable as a general criterion, since it
attributes a special role (and meaning) to the Dwell Time $\tau^{\rm Dw}$,
which on the contrary does {\em not} possess ---in general--- the physical
meaning of global collision--duration, as we showed in Sect.{\bf I-2.}4.\\
\h In the following Section, we are going to show that it is possible to
define (and calculate) ---in a physically meaningful and self-consistent
way--- those durations $\tau_{\rm T}$ \linebreak and $\tau_{\rm R}$.
 
\vspace*{0.75 cm}
 
{\bf I-3.}2 --  {\bf A general definition of the collision durations;
and Applications to Tunnelling}\\
 
\h A direct, general definition of the collision durations was put forth
first by Ohmura$^{[29]}$, and then improved ---and generalized for finite
distances--- by us.$^{[6,8]}$  Following refs.$^{[6]}$, the transmission
and reflection durations $<\tau_{\rm T}>$, $\; <\tau_{\rm R}>$
(averaged over the corresponding flux densities) can be defined, in the
considered case of one--dimensional motion in presence of a barrier, as
follows [T = Traversal]:

\begin{eqnarray}
<\tau_{\rm T}> & \equiv & {<t(x_{\rm f})>}^{\rm III}_{\rm T} -
{<t(x_{\rm i})>}^{\rm I}_{\rm in} = \nonumber\\
 \hfill  \nonumber\\
& = & \frac{\int_{-\infty}^{\infty} {\rm d}t \; t \; J_{\rm T}
^{\rm III}(x_{\rm f},t)}{\int_{-\infty}^{\infty} {\rm d}t \; J_{\rm T}
^{\rm III}(x_{\rm f},t)} \;\; - \;\; \frac{\int_{-\infty}^{\infty} {\rm d}t
\; t \; J_{\rm in}(x_{\rm i},t)}{\int_{-\infty}^{\infty} {\rm d}t \; J_{\rm in}
(x_{\rm i},t)} = \nonumber\\
 \hfill  \nonumber\\
& = & \frac{\int_{0}^{\infty} {\rm d}E \, v \, |gA_{\rm T}|^{2} \;
\tau_{\rm T}^{\rm Ph} (x_{\rm i}, x_{\rm f},\; E)}{\int_{0}^{\infty}
{\rm d}E \, v \, |gA_{\rm T}|^{2}} \equiv (x_{\rm f} - x_{\rm i})<v^{-1}> +
<\Delta \tau_{\rm T}> \; ; \\
 \hfill  \nonumber\\
 \hfill  \nonumber\\
<\tau_{\rm R}> & \equiv & {<t(x_{\rm i})>}^{\rm II}_{\rm R} -
{<t(x_{\rm i})>}_{\rm in} = \nonumber\\
 \hfill  \nonumber\\
& = & \frac{\int_{-\infty}^{\infty} {\rm d}t \; t \; J_{\rm R}
^{\rm II}(x_{\rm i},t)}{\int_{-\infty}^{\infty} {\rm d}t \; J_{\rm R}
^{\rm II}(x_{\rm i},t)} \;\; - \;\; \frac{\int_{-\infty}^{\infty} {\rm d}t
\; t \; J_{\rm in}(x_{\rm i},t)}{\int_{-\infty}^{\infty} {\rm d}t \; J_{\rm in}
(x_{\rm i},t)} = \nonumber\\
 \hfill  \nonumber\\
& = & \frac{\int_{0}^{\infty} {\rm d}E \, v \, |gA_{\rm R}|^{2} \;
\tau_{\rm R}^{\rm Ph} (x_{\rm i}, x_{\rm i},\; E)}{\int_{0}^{\infty}
{\rm d}E \, v \, |gA_{\rm R}|^{2}} \equiv \, 2|x_{\rm i}| \, <v^{-1}> +
<\Delta \tau_{\rm R}> \;\; ,
\end{eqnarray}
 
\
 
which hold when the incoming, reflected and transmitted wave--packets do not
interfere: {\em i.e.}, are totally separated in space--time.  Quantity
$g \equiv g(E-\ove{E})$ was defined in eqs.(5), (6); \ while the ordinary
Phase Times $\tau_{\rm T}^{\rm Ph}, \;\; \tau_{\rm R}^{\rm Ph}$ have been
defined in eqs.(10) and (11). Moreover, quantities $J_{\rm R}^{\rm II}$ and
$J_{\rm T}^{\rm III}$ are defined as in eq.(9), just replacing $\psi (x,k)$
of eq.(5) by $\psi_{\rm R} \equiv A_{\rm R} \; e^{-ikx}$ \ and \
$\psi_{\rm III}
\equiv \psi_{\rm T} \equiv A_{\rm T} \; e^{ikx}$, respectively. \ Let us
stress that our equations (28), (29) do implicitly {\em define} also the time
delays $<\Delta \tau_{\rm T}>, \;\;\, <\Delta \tau_{\rm R}>$ due to
transmission and reflection, respectively; \ as well as the ``average"
instants ${<t(x_{\rm f})>}^{\rm III}_{\rm T}, \;\; {<t(x_{\rm i})>}
^{\rm II}_{\rm R}, \;\; {<t(x_{\rm i})>}_{\rm in}$ at which the corresponding
wave--packets [transmitted, reflected and initial, respectively] pass through
point $x_{\rm f}$ or $x_{\rm i}$. \\
\h Notice that for quasi--monochromatic wave packets, {\em i.e.}, when
approximation (8) holds, eqs.(28), (29) do  directly yield the ordinary
Phase--Times
$\tau_{\rm T}^{\rm Ph}$, $\; \tau_{\rm R}^{\rm Ph}, \;$  given in eqs.(10),
(11).\\
\h However, when $x_{\rm i}, \;\; x_{\rm f}$ are not far from the barrier,
then it happens that the incoming, reflected and transmitted wave--packets
{\em can} interfere.  Moreover, the flux density $J(x,t)$ does in general
change its sign with time; for example, the sign of $J(0,t)$ does change from
$+$ into $-$ approximately a time \ $\hbar \; {\rm d}(\arg A_{\rm R})
/{\rm d}E$ \ after the arrival at $x=0$ of the initial wave--packet.
Therefore, the integrals $\int_{-\infty}^{\infty} {\rm d}t \; t \; J(x,t)$
do represent in general the algebraic sum of positive and negative
quantities, so that the probability densities
$$\frac{J(x,t) \; {\rm d}t}
{\int_{-\infty}^{\infty} {\rm d}t \; J(x,t)}$$
are not positive definite and do not possess a direct physical sense.\\
\h Each probability density acquires a physical meaning only during those
(partial) time--intervals in which the corresponding flux--density $J(x,t)$
does {\em not} change its direction. As a consequence, the previous integrals
are to be split into various integrals, each one carried over a partial
time--interval such that during it
the sign of $J(x,t)$ is only positive, or only negative. Afterwards, one will
sum over all such contributions.\\
\h In other words, we have to deal only with the {\em positive definite}
probability densities $$\frac{{\rm d}t \; J_{+}(x,t)}
{\int_{-\infty}^{\infty} {\rm d}t \; J_{+}(x,t)} \;\;\;\;\; {\rm and} \;\;\;\;\;
\frac{{\rm d}t \; J_{-}(x,t)}
{\int_{-\infty}^{\infty} {\rm d}t \; J_{-}(x,t)} \;\; ,$$  where $J_{+}$ and
$J_{-}$ represent the positive and negative values of $J(x,t)$,
respectively.\\
\h Therefore, we do {\em propose} as physically adequate definitions
for the average {\em transmission time} and the average {\em reflection time}
the following expressions:
 
\begin{eqnarray}
<\tau_{\rm T}> & \equiv & {<t(x_{\rm f})>}_{+} -
{<t(x_{\rm i})>}_{+} = \nonumber\\
 \hfill  \nonumber\\
& = & \frac{\int_{-\infty}^{\infty} {\rm d}t \; t \; J_{+}
(x_{\rm f},t)}{\int_{-\infty}^{\infty} {\rm d}t \; J_{+}
(x_{\rm f},t)} \;\; - \;\; \frac{\int_{-\infty}^{\infty} {\rm d}t
\; t \; J_{+}(x_{\rm i},t)}{\int_{-\infty}^{\infty} {\rm d}t \; J_{+}
(x_{\rm i},t)} \;\; ; \\
 \hfill  \nonumber\\
 \hfill  \nonumber\\
<\tau_{\rm R}> & \equiv & {<t(x_{\rm i})>}_{-} -
{<t(x_{\rm i})>}_{+} = \nonumber\\
 \hfill  \nonumber\\
& = & \frac{\int_{-\infty}^{\infty} {\rm d}t \; t \; J_{-}
(x_{\rm i},t)}{\int_{-\infty}^{\infty} {\rm d}t \; J_{-}
(x_{\rm i},t)} \;\; - \;\; \frac{\int_{-\infty}^{\infty} {\rm d}t
\; t \; J_{+}(x_{\rm i},t)}{\int_{-\infty}^{\infty} {\rm d}t \; J_{+}
(x_{\rm i},t)} \;\; .
\end{eqnarray}
 
\
 
\h Let us notice that, when $x_{\rm f} \geq a$ and $x_{\rm i} \rightarrow
- \infty$, \ equation (30) goes into equation (28) since in that case
$J_{+}(x_{\rm f},t) = J_{\rm T}(x_{\rm f},t) \equiv
J(x_{\rm f},t)$ \ and \ $J_{+}(x_{\rm i},t) = J_{\rm in}(x_{\rm i},t)$.
Analogously, when $x_{\rm i} \rightarrow - \infty$, \ equation (31) goes
into equation (29) since in such a case $J_{+}(x_{\rm i},t) = J_{\rm in}
(x_{\rm i},t)$ \ and \ $J_{-}(x_{\rm i},t) = J_{\rm R}(x_{\rm i},t)$.\\
 
\
 
\h What precedes, and in particular eqs.(30), (31), lead us to adopt as
suitable, strict definitions for the very {\em Tunnelling Time} and the
Reflection Time at the barrier--front (or {\em To-and-Fro Time}) the
following ones:
 
\begin{eqnarray}
<\tau_{\rm tun}> & \equiv & <t(a)>_{+} - <t(0)>_{+} \;\; , \\
 \hfill  \nonumber\\
<\tau_{\rm to-fro}> & \equiv & <\tau_{\rm R}(x_{\rm i} = 0)> \;\; \equiv
\;\; <t(0)>_{-} - <t(0)>_{+} \;\; ,
\end{eqnarray}
 
\
 
where one should recall that the barrier starts at the point $x = 0$.\\
\h According to us, eqs.(32) and (33) are the correct definitions for the
``Tunnelling time" $<\tau_{\rm tun}>$ \ and the
``Reflection-due-to-the-whole-barrier \ time"
({\em i.e.}, the Reflection time at the barrier front wall)
$<\tau_{\rm to-fro}>$. In conclusion, at variance with the authors of
review$^{[1]}$, we think that a positive answer {\em can} be given to their
question about the possibility of a precise, meaningful, univocal
definition of the Tunnelling and Reflection times; such an answer
being provided by our equations (30)--(33).\\
\h Unfortunately, simple analytical expressions for those time--durations
{\em in the energy representation} exist only in particular, limiting
cases. In general, even for Gaussian or quasi--monochromatic wave packets,
calculations can be performed only numerically. Anyway, eqs.(30)--(33)
can be qualitatively tested in an easy way.\\
\h We are left with the question of the time evolution of wave--packets
{\bf inside the barrier}: a problem which till now was paid attention to only
in ref.$^{[7]}$. We shall examine it in the coming Section.\\
\h Before going on, let here mention ---however--- that
\h In the second part of this paper we shall show that our definition 
$<\tau_{\rm T}>$ for the average transmission time results to constitute 
an {\em improvement} with respect to the ordinary dwell--time 
${\ove \tau}^{\rm Dw}$ formula.
 
\vspace*{0.75 cm}
 
{\bf I-3.}3 --  {\bf Time evolution of the tunnelling wave--packets {\em inside}
the barrier}\\
 
\h In ref.$^{[7]}$ calculations were performed of $\rho (x,t)$ and $J(x,t)$,
at different points $x$ inside the barrier, for a Gaussian wave--packet
with an energy spread $\Delta E = 0.025 \; E$. \ The results of those
calculations are presented in Fig.I-3, for $E = \frac{1}{2}V_{0}$ and
$\kappa a = 5/\sqrt{2}$. \ From it, one can see that the times
 $\tau_{\rho}(x), \;\; \tau_{J+}(x)$ and $\tau_{J-}(x)$, taken by the
maximum of $\rho (x,t), \;\; J_{+}(x,t)$ and $|J_{-}(x,t)|$, respectively,
to penetrate the barrier
till the depth $\Delta x = x$, do {\em not} depend {\em linearly} on $x$;
and that $J(x,t)$ ---inside the barrier--- {\em does change} its sign with
time, not very far from the barrier forward wall ($0 \leq x < 0.6 \; a$).\\
\h It is worthwhile to notice that: \ (i) although the continuity equation
$\partial \rho / \partial t + \partial J / \partial x = 0$ \ goes on holding
inside the barrier, nevertheless the equality $J = v\rho$ (which is valid for
quasi-monochromatic wave packets outside the barrier) is {\em not}
valid ---not even approximately--- inside the barrier; and that: \ (ii)
the effective velocities $v_{\rho} \equiv ({\rm d}\tau_{\rho}/{\rm d}x)^{-1},
\;\;\; v_{J+} \equiv ({\rm d}\tau_{J+}/{\rm d}x)^{-1}$ \ and  \ $v_{J-}
\equiv ({\rm d}\tau_{J-}/{\rm d}x)^{-1}$ \ of the maximum of
$\rho, \;\; J_{+}$ and $|J_{-}|$, respectively, not only are
{\em non-constant} as $x$ varies,
but also {\em do not coincide} with each other.\\
\h Passing to the {\em mean velocity} of the wave--packet while tunnelling
through {\em the whole} barrier, it can be defined in a natural way as follows:
 
\begin{equation}
\ove{v_{\rm tun}} \;\; \equiv \;\; \frac{a}{<\tau_{\rm tun}>} \; .
\end{equation}
 
\
 
\h Let us explicitly notice that, if $<\tau_{\rm tun}>$ does not increase with
$a$, the ``effective" speed $\ove{v_{\rm tun}}$ may become arbitrarily large.
This would actually happen when the tunnelling time can be expressed by the
ordinary Phase Time: \ $<\tau_{\rm tun}> = \tau_{\rm T}^{\rm Ph}(0,a;E)$; \
cf. Sect.{\bf I-2.}3.\\
\h Moreover, we can show that even {\em in general}, for
rectangular barriers and large values of $a$, quantity
$<\tau_{\rm tun}>$ does not depend practically on $a$.
In fact, it is obviously \ $<t(a)>^{\rm III}_{\rm T} \; = \; <t(a)>_{+}$; \
so that
the corresponding terms, for $x_{\rm f} = a$, become equal in eq.(32) and in
eqs.(28). Therefore, the same will happen ---for quasi-monochromatic
wave packets--- for the corresponding term included in eqs.(12).  As a
consequence, the term $<t(a)>_{+}$ does not depend on $a$ for {\em opaque}
barriers. \  As to the second term, $<t(0)>_{+}$ , \ of eq.(32), it differs
from \ $<t(0)>_{\rm in}$ \ of eqs.(28) \ [owing to the effect of interference
between
incoming and reflected waves in the flux $\, J_{+}(x_{\rm i}=0, \, t) \,$]
$\;$ by a
quantity that depends on the reflection time \ $\tau^{\rm Ph}_{\rm R}(0,a;E)$,
\ on the wave packet time--extension [of the order of $\hbar/\Delta E$], and
on the form of the wave--packet.  Consequently, also such a ``second" term does
not depend on $a$ for opaque barriers. {\bf We can conclude that the
Hartmann--Fletcher's effect is valid even for our definition (32), so
that our approach by wave-packets confirms that Q.M. predicts Superluminal
tunnelling through opaque barriers}.\\
\h Let us mention, at this point, that another example of barrier ---the 
inverted oscillator potential--- was carefully investigated by Barton,${[30]}$
through a slightly different formalism. In that paper some interesting new
results (which are partially similar to our ones) have been put forth 
in connection with the time evolution of the tunnelling
wave--packets.  In particular, Barton
met cases in which, under the barrier, wave--groups with lower energy travel
faster [cf. also our eq.(28), as well as eq.(12)].
 
\
 
\h It is perhaps worthwhile to add the following observations.  The arriving,
initial wave--packet does interfere with the reflected waves, that start to be
generated as soon as the packet forward tail reaches the barrier edge; in such
a way that (when considering the profiles of fluxes $J(x,t)$ before the
barrier)  the backward tail of
$J_{\rm in}$ decreases ---for destructive interference with $J_{\rm R}$---
in a larger degree than the forward one. This simulates an increase of the
average speed of the entering--flux profile, \  $J_{+}(x,t)$. \ Hence, the
term \ $<t(x)>^{\rm I}_{+}$ \ decreases for negative $x \approx 0$. In other
words, the effective (average) flight--time of the approaching packet from
the source to the barrier does decrease. \ \  Let us
now consider what happens inside the barrier, for positive $x \approx 0$.
An analogous interference effect leads to expect an {\em increase} of the
(effective) tunnelling time $<\tau_{\rm tun}>$; which consequently will not
coincide with the Phase Time \ $\tau^{\rm Ph}_{\rm T}(0,a;E)$: \ not even for
quasi-monochromatic packets.  \ \ Finally, it is interesting to note, and
easy to recognize, that the time--flight decrease before the barrier, and the
tunnelling--time increase inside the barrier, do exactly cancel each other out,
so that the total effect vanishes. \ \ In any case, let us stress that the
fact that the entrance time--instant \ $<t(0)>_{+}$ \ is decreased by the
mentioned ``distortion" does {\em not} obscure the physical sense of our
definition of the tunnelling time \ $<\tau_{\rm tun}>$, \ eq.(32).\\
 
\
 
\h Coming back to the time--evolution of the wave packet inside the barrier,
we cannot describe it as the quasi--classical motion of a particle. We can
visually describe it, on the contrary, in terms of the ``motions" of the
three densities
$\rho^{\rm II}(x,t), \;\; J^{\rm II}_{+}(x,t)$ and $J^{\rm II}_{-}(x,t)$.\\
\h In particular, we can pictorially say that both profiles
[bumps] of the
``incoming"  and the ``reflected" flux--densities $J_{+}(0,t), \;\;
J_{-}(0,t)$ \ (depicted for $x = 0$ as a function of time: see Fig.I-3) \  do
repeat themselves
---with distortion--- for increasing values of the penetration depth $x$.
Such a ``transmission"  takes place rather rapidly (when considering the
velocities $v_{J+}, \;\; v_{J-}$ of the bump maxima) in comparison with
the initial speed $v$, non-uniformly, and
with a gradual decreasing of the bumps [the {\em non-uniformity} of such 
``motions" being
immediately evident from the {\em non-linear} dependence on $x$ of the
stationary wave-function phases inside the barrier: cf. eqs.(1)]. \   In 
particular, such
decreasing is so strong for the reflected waves, that they {\em disappear} at
the end of the barrier. \ At last, we can qualitatively say that the
probability density
$\rho (x,t)$ does conserve the form of its temporal shape at every
position $x$ inside the barrier, but does exponentially decreases as the
depths increases, and quickly, non-uniformly moves towards the final
barrier wall.\\
 
\

\ 

{\bf I-4 -- CONCLUSIONS OF PART I AND PERSPECTIVES}
 
\vspace*{0.75 cm}
 
\h In the Fist Part of this work, besides a critical review of the previous 
definitions of
the tunnelling times and of their consequences, we proposed for them 
---on the basis of our general formalism$^{[6,8]}$--- {\em new}
definitions, eqs.(30)--(33), that we regard as physically acceptable. \ We
do share with the authors of review$^{[1]}$, however, the opinion that
none of the previously known definitions were generally acceptable.\\
\h In the First Part above we began analyzing ---for the first time--- also
the time--evolution of the tunnelling process {\bf inside} the barrier.\\
\h Let us recall that our formalism is based on the introduction and 
application of a [non--selfadjoint, but {\em hermitian}] operator for 
Time as well as on suitable definitions for the observables' time--averages 
[cf. eqs.(28)--(31)]. The physical self-consistency of such a formalism
may be regarded as supported by results as: (i) the validity of a 
correspondence principle
between the time--energy QM commutation relation and the CM Poisson brackets;
 \ (ii) the validity of an Ehrenfest principle for the average 
time--durations;$^{[6,8]}$
\ (iii) the coincidence of the quasi-classical limit of our own QM definitions
for time durations (when such a limit exists: {\em i.e.}, for above--barrier
energies) with the analogous, well-known expressions of classical mechanics
[see in particular the second and third refs.$^{[6]}$, and refs. therein]. \ 
Moreover, definitions
similar to eqs.(28)--(29) have been already applied and tested in the analysis
of experimental data on nuclear reaction durations, in the range
$10^{-21}-10^{-15}$s, obtained by blocking--effect, X--ray spectroscopy and 
brehmsstrahlung experiments [see in particular the first two refs.$^{[6]}$, 
and refs. therein].
 \ Let us stress that for completely extracting the time--duration values from
experimental data on the abovementioned (non--stationary) processes, which are
always non--linearly depending on the nuclear reaction durations, it is 
necessary to have recourse not only to equations of the type (28)--(29), but 
also to correct definitions for the duration variances, and for the 
duration distribution higher--order central moments:$^{[6]}$ which is provided 
by our formalism. \ At last, let us mention that such a formalism did provide 
also useful tools for the resolution of  some long--standing
problems related to the time--energy uncertainty relation.${[6,8]}$\\

\h As to the future, let us before all recall that analytical expressions
in the energy representation for $<\tau_{\rm tun}>$ and $<\tau_{\rm to-fro}>$
do not exist, even in the case of the simplest barriers. However, numerical
calculations can be straightforwardly performed about the time evolution of
the considered wave--packet inside the barrier.  On the basis of such
one-particle (and one-dimensional) calculations, it will be possible to
start developing a {\em kinematical theory} for the tunnelling of bound or
metastable many--particle systems, and of unbound aggregates, through
various barriers.\\
\h Within the field of nuclear physics, it will be interesting ---on the
basis of the peculiarities of the tunnelling--process temporal evolution
(Sect.{\bf I-3.}3 above; and ref.$^{[7]}$)--- to investigate the possibility
of {\em observing} effects due to the change in form, in volume, in orientation
and in life--time of many--particle systems (nuclei, fragments,...) during their
tunnelling. As well as effects due to ``collisions", inside the barrier
({\em e.g.}, the ion--ion barrier), between different, successive, penetrating
particles: for example, the role of those possible collisions in enhancing
two--proton sub-barrier transfer in heavy ion interactions. \ Another future
task will be developing a multi-dimensional description, in terms of
tunnelling processes, of the sub-barrier fusion of two nuclei: taking account
of nuclear deformation, formation of noses or a neck, and of dissipation
phenomena.$^{[31]}$\\
\h At last, we should comment on the fact that ---when \ $<\tau_{\rm tun}>$ \
does not practically depend on the barrier width $a$, as we have seen to
happen for opaque or high barriers (Sects.{\bf I-3.}3 and {\bf I-2.}3)---
then one apparently meets, in connection with eq.(34), speeds that ({\em
inside the barrier}) can assume arbitrarily large values.  This does not
violate any postulate, {\em as far as} we deal with non-relativistic (quantum)
physics; and in fact such a phenomenon has already been frequently met within
``quantum systems".$^{[32]}$\\
\h It is easy to check that the same would happen, in our case,  {\em even}
when replacing the
Schroedinger equation by the Klein--Gordon or Dirac equations, at least for
barriers that do not depend explicitly on time. More interesting is the
occurrence of this fact in such ``quantum field theories" (which deal,
however, with semi--classical potentials).\\  
\h One could think that those 
infinitely large speeds might disappear in a {\em self-consistent} 
relativistic quantum theoretical treatment. But such an expectation seems
to be wrong, for reasons that come also from Special Relativity itself.[33] 
The Hartman (Superluminal tunnelling) effect should then be added
to the already known results that suggest the existence of Superluminal
motions: cf., {\em
e.g.}, refs.$^{[34]}$. Results that have been predicted, inside relativistic
theories like QFT, by many authors, like Sudarshan$^{[35]}$, Van Dam and
Wigner$^{[36]}$, Recami$^{[33]}$, Ne'eman$^{[37]}$ and others.
 
\vspace*{1.5 cm}

{\bf I-5 -- Note added in the 1992 proofs:}
 
\vspace*{0.75 cm}

In two recent papers$^{[38,39]}$, B\"{u}ttiker and Landauer published some other
critical comments about the various approaches examined by us in Sect.{\bf I-2}.
Let us mention, among them, a further criticism  of Hauge and Stovneng's
relationship [eq.(27), Sect.{\bf I-3}.1], appeared in ref.$^{[39]}$. Moreover, an
interesting, brief review of the first experimental measurements of tunnelling
times did appear in ref.$^{[38]}$.\\  

\

\
 
{\bf CAPTIONS OF THE FIGURES OF PART I:}
 
\vspace*{0.75 cm}
 
Fig.I-1 -- The case of {\em stationary} scattering and tunnelling. In this
figure we depict a generic potential--barrier $V(x)$ (which does not
depend explicitly on time), with the three regions generated by its
presence; and the incoming, reflected and transmitted
plane waves.
 
\
 
Fig.I-2 -- Orientations of spin and magnetic field for the case of the
``Larmor clock" (see the text).
 
\
 
Fig.I-3 -- A pictorial view of the time dependence of $J(x,t)$ and $\rho
(x,t)$, for various values of the penetration depth $x$ inside a
rectangular barrier (for Gaussian wave--packets). Actually, as $x$
increases, all the ``bumps" suffer an exponential damping; but, for
convenience, we neglected in this figure the exponential factor
$\exp (- \kappa x)$.\\

\

BIBLIOGRAPHY OF PART I:

\newpage

\centerline{{\large{\bf SECOND PART}}}

\

\

{\bf II-1. -- Introduction}

\vs{0.5 cm}

\h In the First Part$^{[\IIrm-1]}$ of this paper, we have put forth an analysis 
of the main
theoretical definitions of the sub-barrier tunnelling and reflection times,
and proposed new definitions for such durations which seem to be
self-consistent within {\em conventional} quantum mechanics.$^{\# 1}$ 
\footnotetext{$^{\# 1}$ Let us take advantage of the present opportunity for 
pointing out that a misprint entered our eq.(10) in ref.[II-1], whose last
term $ka$ ought to be eliminated. \ Moreover, due to an {\em editorial} error, 
in the footnote at page 32 of our ref.[II-16] the dependence of $G$ on $\Delta k$
disappeared, whilst in that paper we had assumed \ $G(k-\ove{k}) \equiv C \,
\exp [-(k-\ove{k})^2 / (\Delta k)^2]$.}
 \ In particular, the ``prediction" by our
theory$^{[\IIrm-1]}$ of the reality of the {\em Hartman effect\/}$^{[\IIrm-2]}$ in 
tunnelling 
processes has recently received ---due to the analogy$^{[\IIrm-3]}$ between 
tunnelling electrons and evanescent waves--- quite interesting, even if 
indirect,
experimental verifications at Cologne,$^{[\IIrm-4]}$ Berkeley,$^{[\IIrm-5]}$ 
Florence$^{[\IIrm-6]}$ and Vienna.$^{[\IIrm-6]}$

\h Main aims of this Second Part [that appeared in print in J. de Physique-I 
5 (1995) 1351] are: \ presenting and analysing the results of several
numerical calculations of the penetration and return times {\em inside} a 
rectangular potential barrier during tunnelling (Sect.{\bf II-3}); \ and 
proposing new suitable formulae for the
distribution variances of the transmission and reflection times 
(Sect.{\bf II-2}).
\h The results of our numerical evaluations
seem to confirm that our approach is
physically acceptable, and that it moreover 
implied, and implies, the existence of the so-called 
{\em ``Hartman effect\/"} (i.e., of tunnelling with Superluminal 
group-velocities) even for {\em non}--quasi-monochromatic packets.\\ 

\h Before all, let us add here ---however--- some brief comments about a few 
furthet papers, appeared recently:\\ 

\h (i) First, let us mention that in Part I above we have overlooked a new 
expression for the dwell--time ${\ove \tau}^{\rm Dw}$ derived by Jaworsky and 
Wardlaw$^{[\IIrm-7,8]}$ 

\
 
\hfill{$
{\ove \tau}^{\rm Dw}(x_{\rm i},x_{\rm f};k) \; = \; \left(
{\int_{-\infty}^{\infty} {\rm d}t \, t \, J(x_{\rm f},t) \: - \:
\int_{-\infty}^{\infty} {\rm d}t \, t \, J(x_{\rm i},t)}\right)
\left({\int_{-\infty}^{\infty} {\rm d}t \; J_{\rm in}(x_{\rm i},t)}
\right)^{-1} \;\; ,$
\hfill} (II-1)
 
\

which is indeed equivalent$^{[\IIrm-7]}$ to our eq.(16) of ref.[II-1] (all 
notations being defined therein):
 
\   
 
\hfill{$
{\ove \tau}^{\rm Dw}(x_{\rm i},x_{\rm f};k) \; = \; \left(
{\int_{-\infty}^{\infty} {\rm d}t \int_{x_{\rm i}}^{x_{\rm f}} {\rm d}x \;
\rho (x,t)}\right) \left({\int_{-\infty}^{\infty} {\rm d}t \;
J_{\rm in}(x_{\rm i},t)} \right)^{-1} \;\; .$
\hfill} (II-2)
 
\ 

This equivalence
{\em reduces} the difference, between our definition 
$<\tau_{\rm T}>$ of the
average transmission time ---under our assumptions---  and quantity 
${\ove \tau}^{\rm Dw}$, {\em to} the difference
between the average made by using the positive--definite probability
density ${\rm d}t \, J_{+}(x,t)/\int_{-\infty}^{\infty} 
{\rm d}t \, J_{+}(x,t)$
and the average made by using the ordinary ``probability density"
${\rm d}t \, J(x,t)/\int_{-\infty}^{\infty} {\rm d}t \, J(x,t)$. \ Generally
speaking, the last expression is {\em not} always positive definite,
as it was explained at page 350 of ref.[II-1], and hence does not possess
any direct physical meaning.\\ 

\h (ii) In ref.[II-9] an attempt was made to analyze the evolution of the
wave packet mean position $<x(t)>$ (``center of gravity"), averaged over
$\rho {\rm d}x$, during its tunnelling through a potential barrier. \ Let
us here observe that the conclusion to be found therein, about the absence
of a causal relation between the incident space centroid and its 
transmitted equivalent, holds {\em only} when the contribution 
coming from the barrier region to the space integral is negligible.\\  

\h (iii) Let us also add that in ref.[II-10] it was analyzed the 
{\em distribution}
of the transmission time $\tau_{\rm T}$ in a rather sophisticated way, which
is very similar to the dwell--time approach, however with an 
{\em artificial}, abrupt switching on of the initial wave packet. \ \  
We are going to propose, on the contrary, and in analogy with our 
eqs.(30)-(31) in
ref.[II-1], the following expressions, as physically adequate definitions 
for the {\em variances} (or dispersions)
${\rm D} \, \tau_{\rm T}$ and ${\rm D} \, \tau_{\, \rm R}$ of the transmission and
reflection time [see Sect.{\bf II-2}], respectively:

\

\hfill{${\rm D} \, \tau_{\rm T} \; \equiv \; {\rm D} \, t_{+}(x_{\rm f}) + 
{\rm D} \, t_{+}(x_{\rm i})$\hfill} (II-3)

\

and

\

\hfill{${\rm D} \, \tau_{\, \rm R} \; \equiv \; {\rm D} \, t_{-}(x_{\rm i}) + 
{\rm D} \, t_{-}(x_{\rm i}) \; ,$\hfill} (II-4)

where

\vskip 1. mm

\
 
\hfill{$
{\rm D} \, t_{\pm}(x) \; \equiv \; \disp{
\frac{\int_{-\infty}^{\infty} {\rm d}t \; t^2 \, J_{\pm}(x,t)}
{\int_{-\infty}^{\infty} {\rm d}t \, J_{\pm}(x,t)} } \; - \;
\disp{ \left( \frac{\int_{-\infty}^{\infty} {\rm d}t \; t \; J_{\pm}(x,t)}
{\int_{-\infty}^{\infty} {\rm d}t \, J_{\pm}(x,t)} \right) ^2} \; .$
\hfill} (II-5)
 
\

\vskip 1.5 mm

Equations (I-3)--(I-5) are based on the formalism expounded in refs.[II-11],
as well as on our definitions for $J_{\pm}(x,t)$ in ref.[II-1]. \ Of course, we
are supposing that the integrations over \ $J_{+}(x_{\rm f})  \, {\rm d}t$, \
$J_{+}(x_{\rm i})  \, {\rm d}t$ \ and \ $J_{-}(x_{\rm i})  \, {\rm d}t$
are independent of one another. \ We shall devote Sect.{\bf II-2}, below, to 
these problems, i.e., to the problem of suitably defining mean 
values and variances  of durations, for various transmission and reflection
processes during tunnelling.\\ 

\h (iv) Below, in Sect.{\bf II-4}, we shall briefly 
re-analyse some other definitions of tunnelling durations.\\

\h Before going on, let us recall that several reasons ``justify" the 
existence of different approaches to the definition of tunnelling times: \ \
(a) the problem of defining tunnelling durations is  closely connected 
with that of defining a time operator, i.e., of introducing {\em time} as a 
(non-selfadjoint) quantum mechanical observable, and subsequently of 
adopting a general definition for collision durations in
 quantum mechanics. Such preliminary problems did receive some 
 clarification in recent times (see, for
example, ref.[II-1] and citations [8] and [22] therein); \ \ (b) the motion of a particle 
tunnelling inside a potential barrier
is a purely quantum phenomenon, devoid of any classical, intuitive limit; \ \
(c) the various theoretical approaches may differ in the  
choice of the boundary conditions or in the modelling of the experimental
situations.\\

\

{\bf II-2. -- Mean values and Variances for various Penetration 
and Return Times during tunnelling}

\vs{.5 cm}                                             

\h In our previous papers, and in Part I above, we 
proposed for the {\em transmission} and {\em reflection} times some 
formulae which
imply ---as functions of the penetration depth--- integrations over time of
$J_{+}(x,t)$ and $J_{-}(x,t)$, respectively. \ Let us recall that 
the total flux $J(x,t)$ inside a barrier consists of two components, 
$J_{+}$ and $J_{-}$, associated with motion along the positive and the 
negative $x$-direction, respectively. \ Work in similar directions 
did recently appear in ref.[II-12]. 

\h Let us refer ourselves ---here--- to    
tunnelling  and reflection processes of a particle by a potential barrier, 
confining ourselves to one space dimension. \ Namely, let us study 
the evolution of a wave packet
$\Psi(x,t)$,  starting from the initial state $\Psi_{\rm {in}}(x,t)$; 
and follow the notation 
introduced in ref.[II-1]. \ In the case of uni-directional  motions it is  
already known$^{[\IIrm-13]}$ that the 
flux density \ $J(x,t) \equiv {\rm Re}[ (i \hbar / m) \, 
\Psi(x,t) \; 
{\partial \Psi^{*}(x,t) / \partial x}]$ \ can be actually interpreted as the 
probability that the 
particle (wave packet) passes through position $x$ during a unitary 
time--interval 
centered at $t$, {\bf as it easily follows from the continuity equation and 
from the fact that quantity} \ $\rho(x,t) \equiv |\Psi(x,t)|^{2}$ \ {\bf is 
the probability density} for our 
``particle" to be located, at time t, inside a unitary 
space--interval centered at $x$. \ \ Thus, in order to determine  
the {\em mean} instant at which a moving wave packet 
$\Psi(x,t)$ passes through position $x$, we have to take the average
of the time variable $t$ with respect to the weight \ $w(x,t) \equiv
J(x,t)/ \int_{-\infty}^{\infty} J(x,t) \, \drm t$. 
 
\h However, if the motion 
direction can {\em vary}, then quantity $w(x,t)$ is no longer positive 
definite, and moreover is not bounded, because of the variability of 
the $J(x,t)$ sign.
 \ In such a case, one can introduce the two weights:

$$w_{+}(x,t) = J_{+}(x,t) \; \left[ \int_{-\infty}^{\infty} J_{+}(x,t)
 \: \drm t  \right]^{-1} 
\eqno(\IIrm-6)$$ 

\

$$w_{-}(x,t) = J_{-}(x,t) \; \left[ \int_{-\infty}^{\infty} J_{-}(x,t) 
 \: \drm t \right]^{-1}  \; , 
\eqno(\IIrm-7)$$ 
  
\

where $J_{+}(x,t)$ and $J_{-}(x,t)$ represent the positive and negative 
parts of $J(x,t)$, respectively, which are bounded, positive--definite 
functions, normalized to 1.  \ 
Let us show that, from the ordinary probabilistic interpretation of 
$\rho (x,t)$ and from the well-known continuity equation

$$ {\partial{\rho(x,t)} \over {\partial t}} + {\pa J(x,t) \over \pa x} = 0  
\ ,\eqno(\IIrm-8)$$

it  follows {\em also in this (more general) case} that quantities 
$w_+$ and $w_-$,
represented by eqs.(II-6), (II-7), can be regarded as the probabilities that 
our ``particle" passes through position $x$ during 
a unit time--interval centered
at $t$ (in the case of forward and backward motion, respectively).

\h Actually, for those time intervals for which $J = 
J_{+}$ or $J = J_{-}$, one can rewrite eq.(II-8) as follows:

$${{\partial {\rho_{>}(x,t)} \over \partial t} = 
- {\pa J_{+}(x,t) \over {\pa x}}} \eqno(\IIrm-9.a)$$

\

$${{\partial {\rho_{<}(x,t)} \over \partial t} = -{\pa J_{-}(x,t) \over 
\pa x} \ ,}
\eqno(\IIrm-9.b)$$

\           

respectively. \ Relations (9.a) and (9.b) can be considered  as formal
definitions of \ $\partial {\rho_{>}} / \partial t$ and \ $\partial 
{\rho_{<}} / \partial {t}$. \ \ Let us now integrate eqs.(II-9.a), (II-9.b) 
over time from $-\infty$ to $t$; \ we obtain:

$$\rho_{>}(x,t)= -\int_{-\infty}^{t}  
{\pa J_{+}(x,{t}') \over \pa x} \: \drm {t}' \eqno(\IIrm-10.a)$$

$$\rho_{<}(x,t)= -\int_{-\infty}^{t}  
 {\pa J_{-}(x,{t}') \over \pa x} \: \drm t' \eqno(\IIrm-10.b)$$

\ 

with the initial conditions \ $\rho_{>}(x,-\infty)=\rho_{<}(x,-\infty)=0$.
 \ Then, let us introduce the quantities

$$N_{>} (x,\infty;t) \equiv \int_{x}^{\infty} \rho_{>}({x}',t) \, \drm {x}' 
= \int_{-\infty}^{t} J_{+}(x,{t'}) \, \drm {t}' \ >0 \eqno(\IIrm-11.a)$$

\
 
$$N_{<} (-\infty,x;t) \equiv \int_{-\infty}^{x} \rho_{<}({x}',t) \, 
\drm {x}' =
-\int_{-\infty}^{t} J_{-}(x,{t}') \, \drm {t}' \ >0 \ , \eqno(\IIrm-11.b)$$

\

which have  the meaning of probabilities for our ``particle" to be located  
at time $t$ on the semi-axis  $(x,\infty)$ or $(-\infty,x)$ 
respectively, as functions of
the flux densities $J_{+}(x,t)$ or  $J_{-}(x,t)$,
provided that the normalization condition \
$\int_{-\infty}^{\infty}\rho(x,t) \drm x = 1$ \ is fulfilled.  \
The r.h.s.'s of eqs.(II-11.a) and (II-11.b) have been obtained by integrating the
r.h.s.'s of eqs.(II-10.a) and (II-10.b) and adopting the boundary conditions  \
$J_{+}(-\infty,t) = J_{-}(-\infty,t) = 0$. \  Now, by
differenciating eqs.(II-11.a) and (II-11.b) with respect to $t$, one obtains:

\
$${{\partial{N_{>}}(x,\infty,t) \over \partial{t}} = 
J_{+}(x,t)  > 0 } \eqno(\IIrm-12.a)$$

\

$${{\partial{N_{<}}(x,-\infty,t) \over \partial{t}} = 
- \, J_{-}(x,t)  > 0 } \ . \eqno(\IIrm-12.b)$$

{\em Finally}, from eqs.(II-11.a), (II-11.b), (II-12.a) and (II-12.b), one can 
infer that:

$${w_{+}(x,t) ={{\partial {N_{>}}(x,\infty;t)/\partial {t} 
\over {N_{>}(x,-\infty;\infty)}}}} \eqno(\IIrm-13.a)$$

$${w_{-}(x,t) ={{\partial {N_{<}}(x,-\infty;t)/\partial {t} 
\over {N_{<}(-\infty,x;\infty)}}}} \ , \eqno(\IIrm-13.b)$$

which justify the abovementioned probabilistic interpretation of $w_{+}(x,t)$ 
and $w_{-}(x,t)$. \ Let us notice, incidentally, that our approach does 
{\em not} assume any ad hoc postulate, contrarily to what believed by the
author of ref.[II-14].\\

\h At this point, we can eventually define the {\em mean value} of the time 
at which our ``particle"  passes
through position $x$,  travelling in the positive or negative direction of the
$x$ axis, respectively, as:

\
$$<t_{+}(x)> \ \equiv \ {{\int_{-\infty}^{\infty} t \, J_{+}(x,t) \, 
\drm t  \over \int_{-\infty}^{\infty} J_{+}(x,t) \, \drm t }} \eqno(\IIrm-14.a)$$

\
$$<t_{-}(x)> \ \equiv \ {{\int_{-\infty}^{\infty} \, t J_{-}(x,t) \, 
\drm t  \over \int_{-\infty}^{\infty} J_{-}(x,t) \, \drm t }} \eqno(\IIrm-14.b)$$

and, moreover, the {\em variances} of the distributions of these times as:

$${\rm D} \, t_+(x) \ \equiv \ {{\int_{-\infty}^{\infty} t^{2} J_{+}(x,t) \drm t  \over
\int_{-\infty}^{\infty} J_{+}(x,t) \drm t }} - [<t_+(x)>]^{2} \eqno(\IIrm-15.a)$$

\

$${\rm D} \, t_-(x) \  \equiv \ {{\int_{-\infty}^{\infty} t^{2} J_{-}(x,t) \drm t  \over
\int_{-\infty}^{\infty} J_{-}(x,t) \drm t }} - [<t_-(x)>]^{2} \ , 
\eqno(\IIrm-15.b)$$

in accordance with the proposal presented in refs.[II-1,15].

\h Thus, we have a formalism for defining mean values, variances
(and other central moments) related to the duration {\em distributions}
of all possible processes for a particle, tunnelling
through a potential barrier located in the interval $(0,a)$
along the $x$ axis; and not only for tunnelling, but also for all possible 
kinds of collisions, with arbitrary energies and potentials.  \
For instance, we have that

$$<\tau_{\rm T}(x_{\rm i},x_{\rm f})> \ \equiv \ <t_+(x_{\rm f})> - 
<t_+(x_{\rm i})> \eqno(\IIrm-16)$$

 \
 
with $-\infty < x_{\rm i} < 0$ \ and \ $a < x_{\rm f} < \infty$; \ and 
therefore \ (as anticipated in eq.(II-3)) \ that 

$${\rm D} \; 
\tau_{\rm T}(x_{\rm i},x_{\rm f}) \; \equiv \; {\rm D} \, t_{+}(x_{\rm f}) + 
{\rm D} \, t_{+}(x_{\rm i}) \ ,$$ 

for {\em transmissions} from 
region $(-\infty,0)$ to region $(a,\infty)$ which we called$^{[\IIrm-1]}$ regions I 
and III, respectively. \ \ Analogously, for the pure (complete) tunnelling 
process one has:

 \
$$<\tau_{\rm Tun}(0,a)> \ \equiv \ <t_+(a)> - <t_+(0)> \eqno(\IIrm-17)$$

and

$${\rm D} \; \tau_{\rm Tun}(0,a) \; \equiv \; {\rm D} \, t_{+}(a) + {\rm D} \, 
t_{+}(0) \ ; \eqno(\IIrm-18)$$
 
\

while one has

$$<\tau_{\, \rm Pen} (0,x_{\rm f})> \ \equiv \ <t_+(x_{\rm f})> - <t_+(0)> 
\eqno(\IIrm-19)$$

\
 
and 

$${\rm D} \; \tau_{\, \rm Pen} (0,x_{\rm f}) \ \equiv \  
{\rm D} \, t_+(x_{\rm f}) + {\rm D} \, t_+(0) 
\eqno(\IIrm-20)$$

\
  
[with \ $0<x_{\rm f}<a$] \ for {\em penetration} inside the barrier region 
(which we called region II).  \ \ Moreover:

$$<\tau_{\, \rm Ret}(x,x)> \ \equiv \ <t_-(x)> -<t_+(x)> \eqno(\IIrm-21)$$

$${\rm D} \; \tau_{\, \rm Ret}(x,x) \ \equiv \ {\rm D} \, t_-(x) 
 + {\rm D} \, t_+(x) \eqno(\IIrm-22)$$

[with \ $0<x<a$] \ for ``{\em return processes\/}"  inside the barrier.  \ At
last, for {\em reflections} in region I, we have that:

$$<\tau_{\, \rm R}(x_{\rm i},x_{\rm i})> \ \equiv \  <t_-(x_{\rm i})> - 
<t_+(x_{\rm i})> \eqno(\IIrm-23)$$
 
[with $-\infty<x_{\rm i}<a$], \ and \ (as anticipated in eq.(II-4)) \ that \ 
${\rm D} \; \tau_{\, \rm R}(x_{\rm i},x_{\rm i}) \ \equiv \ 
{\rm D} \, t_-(x_{\rm i}) + {\rm D} \, t_+(x_{\rm i})$. 

\h Let us stress that our definitions hold within
the framework of conventional quantum mechanics, without the introduction 
of any new postulates, and with the single measure expressed by weights
$(13.a), \ (13.b)$ for all time averages (both in the initial and in the
final conditions).

\h According to our definition, the tunnelling phase time 
(or, rather, the transmission duration), defined by the stationary phase
approximation for quasi-monochromatic particles, is meaningful {\em only}
in the limit $x_{\rm i} \rightarrow \infty$ when $J_{+}(x,t)$ is the flux density
of the initial packet $J_{\rm in}$ of {\em incoming} waves  ({\em in absence 
of reflected waves}, therefore).
 
\h Analogously, the dwell time, which can be represented (cf. eqs.(II-1),(II-2)) 
by the expression$^{[\IIrm-7,8,16]}$

$${{{\ove \tau}^{\rm Dw}(x_{\rm i},x_{\rm f}) = \left[ 
\int_{-\infty}^{\infty} t \; J(x_{\rm f},t) \; \drm t - 
\int_{-\infty}^{\infty} t \; J(x_{\rm i},t) \; \drm t \; \right] \; \left[
\int_{-\infty}^{\infty} J_{\rm in}(x_{\rm i},t) \; \drm t \,
\right]^{-1} }} \ ,$$

\

with $-\infty < x_{\rm i} < 0$, and $x_{\rm f}>a$, \ is not  
acceptable, generally speaking.  In fact, 
the weight in the time averages is meaningful, positive definite 
and normalized to 1 {\em only} in the rare cases when
$x_{\rm i} \longrightarrow-\infty$ and $J_{\rm in} = J_{\rm III}$ \ (i.e., 
when the barrier is transparent).

\vspace*{1. cm}

{\bf II-3. -- Penetration and Return process durations, inside a rectangular
barrier, for tunnelling gaussian wave packets: Numerical results}

\vspace*{0.5 cm}

\h We put forth here the results of our calculations of mean durations for  
various penetration (and return) processes, {\em inside} a rectangular 
barrier, for tunnelling gaussian wave packets;  
one of our aims being to investigate the 
{\em tunnelling speeds}. \ In our calculations, the initial wave packet is

$$\Psi_{\rm in}(x,t) = \int_{0}^{\infty}G(k-\overline k) \; 
\exp[ikx-iEt/\hbar] \; \drm k \eqno(\IIrm-24)$$ 
 
\

with  

$$G(k- \overline k) \equiv C \exp [{- (k-\overline k)^2} / 
{(2 \, \Delta k)^{2}}] \ , \eqno(\IIrm-25)$$

{\em exactly} as in ref.[II-8]; \ and with \ 
$E = \hbar^{2} k^{2}/2m$; \ quantity
$C$ being the normalization constant, and $m$ the electron mass. \ 
Our procedure  of integration was described 
in ref.[II-16]. 

\

\h Let us express the penetration depth in {\aa}ngstroms, and
the penetration time in seconds. \ \ In Fig.II-1  we show the plots 
corresponding to \ $a = 5 \;$\AA, \ for \ $\Delta k = 0.02$ and 
$0.01 \; {\rm {\AA}}^{-1}$, \ respectively. \ The penetration time 
$<\tau_{\, \rm Pen}>$ always tends to a {\em saturation} value.\\ 
\h In Fig.II-2 we show,
for the case \ $\Delta k = 0.01 \; {\rm {\AA}}^{-1}$, \ the plot corresponding 
to \ $a = 10 \;$\AA. \ It is interesting that $<\tau_{\, \rm Pen}>$ is 
practically {\em the same} (for the same $\Delta k$) for \ $a = 5$ and 
$a = 10 \;$\AA, a result that confirm, let us repeat, the existence$^{[\IIrm-1]}$ 
of the so-called 
{\em Hartman  effect\/}.$^{[\IIrm-2]}$ \ \ Let us add that, \ when varying the 
parameter $\Delta k$ between $0.005$
and $0.15 \; {\rm \AA}^{-1}$ and letting $a$ to assume values even larger than 
$10 \; {\rm \AA}$, \ analogous results have been always gotten. \ Similar 
calculations have been performed (with
quite reasonable results) also for various energies $\ove{E}$ in the 
range $1$ to $10$ eV.~$^{\# 2}$
\footnotetext{$^{\# 2}$ For the interested reader, let us recall that, when  
integrating over d$t$, we used the interval \ $- 10^{-13} \:$s \ to \ 
$+10^{-13} \:$s \ (symmetrical 
with respect to $t = 0$), \ very much larger than the temporal wave packet 
extension. [Recall that the extension in time of a wave packets is of the
order of \ $1/(\ove{v} \, 
{\Delta k}) = ({\Delta k} \, \sqrt{2 \ove{E} / m})^{-1} \simeq 10^{-16} \:$s]. 
\ Our ``centroid" has been always $t_{0} = 0; \; x_{0} = 0$. \ \  For 
clarity's 
sake, let us underline again that in our approach the initial
wave packet $\Psi_{\rm in} (x,t)$ is not regarded as prepared at a certain
instant of time, but it is expected to flow through any (initial) point 
$x_{\rm i}$ during the infinite time interval ($-\infty, \;\; +\infty$),
even if with a {\em finite} time--centroid $t_0$. \ The value of such
centroid $t_0$ is essentially defined by the phase of the weight amplitude 
$G(k- \ove{k})$, and in our case is equal to 0 when $G(k- \ove{k})$ is 
real.}

\h In Figs.II-3, 4 and 5 we show the behaviour of the mean penetration 
and return durations as function of the penetration depth 
(with $x_{\rm i} = 0$ and
$0\le \, x_{\rm f} \equiv x \, \le a$), \ for barriers with height $V_{0} = 10$ eV
and width $a= 5 \;${\AA} or $10 \;$\AA. \ \ 
In Fig.II-3 we present the plots of 
$<\tau_{\, \rm Pen}(0,x)>$ corresponding to different values of the mean
kinetic energy: \  $\overline E$ = 2.5 eV, $\;$5 eV and 7.5 eV (plots 1, 2
and 3, respectively) with $ \Delta k=0.02 {\rm \AA}$; \  and 
$\overline E = 5$ eV with $\Delta k =0.04 {\rm \AA}^{-1}$ (plot 4), 
always with $a= 5 {\rm \AA}$. \ \
In Fig.II-4 we show the plots of $<\tau_{\, \rm Pen}(0,x)>$,
corresponding to $a=5 {\rm \AA}$, with $\Delta k = 0.02 {\rm \AA}^{-1}$ 
and 0.04 ${\rm \AA}^{-1}$
(plots 1 and 2, respectively); and to $a=10 {\rm \AA}$, 
with $\Delta k = 0.02 {\rm \AA}^{-1}$
and 0.04 ${\rm \AA}^{-1}$ (plots 3 and 4, respectively), the mean 
kinetic energy $\overline E$ being 5 eV, i.e., one half of $V_{0}$. \ \ 
In Fig.II-5 the plots are shown 
of $<\tau_{\, \rm Ret}(x,x)>$. \ The curves 1, 2 and 3 correspond to 
$\overline E = 2.5$ eV, $\;$5 eV and 7.5 eV, respectively, for $\Delta k =
0.02 {\rm \AA}^{-1}$ and  $a = 5  {\rm \AA}$; \ the curves 4, 5 and 6 
correspond to $\overline E = 2.5$ eV, $\;$5 eV  and 7.5 eV, respectively, for
$\Delta k = 0.04 {\rm \AA}^{-1}$ and $a = 5 {\rm \AA}$; \ while the curves 
7, 8 and 9
correspond to $\Delta k =0.02 {\rm \AA}^{-1}$ and 0.04 ${\rm \AA}^{-1}$,
respectively, for $\overline E = 5$ eV and $a=10 {\rm \AA}$.
  
\h Also from the new Figs.II-3--5 one can see that: \ 1) at variance with 
ref.[II-8], no plot considered by us for the mean penetration duration 
$<\tau_{\, \rm Pen}(0,x)>$ of our wave packets presents  any interval 
with negative values, 
nor with a decreasing  $<\tau_{\, \rm Pen}(0,x)>$
for increasing  $x$; \ and, moreover, that \  2) the mean tunnelling 
duration
$<\tau_{\rm Tun}(0,a)>$  does not depend on the barrier  width $a$
(``Hartman effect"); \ and finally that \ 3) quantity 
$<\tau_{\rm Tun}(0,a)>$ decreases when the energy increases. \ \ Furthermore,
it is noticeable that also from Figs.II-3--5 we observe: \ 4) a rapid increase 
for the value of the electron penetration time in the
initial part of the barrier region (near $x= 0$); \ and \ 5) a tendency of
$<\tau_{\, \rm Pen}(0,x)>$ to a saturation value in the final part of 
the barrier, near $x=a$.

\h Feature 2), firstly 
observed for quasi-monochromatic particles,$^{[\IIrm-2]}$
does evidently agree with the predictions made in ref.[II-1] for 
arbitrary wave packets. \ Feature 3) is also in agreement with previous 
evaluations performed
for quasi-monochromatic particles and presented, for instance, in 
refs.[II-1,2,15]. \ Features 4) and 5) can be apparently explained by 
interference between 
those initial penetrating and returning waves inside 
the barrier, whose superposition yields the resulting fluxes $J_{+}$ and
$J_{-}$. In particular, if in the initial part of the barrier 
the returning--wave packet is comparatively large, it
does essentially extinguish the leading edge 
of the incoming--wave packet. \ By contrast, if for growing $x$
the returning--wave packet quickly vanishes, then the contribution 
of the leading edge of the incoming--wave packet to the mean penetration 
duration $<\tau_{\, \rm Pen} (0,x)>$
does initially (quickly) grow, while in the final barrier region its increase  
does rapidly slow down. 
    
\h Furthermore, the larger is the barrier width $a$, the larger is the part of the 
back edge of the incoming--wave packet which is
extinguished by interference with the returning--wave packet. \ 
Quantitatively,
these phenomena will be studied elsewhere. \ Finally, in connection with 
the plots 
of $<\tau_{\, \rm Ret}(x,x)>$ as a function of $x$, presented in Fig.II-5,
let us observe that: \ (i) the mean reflection duration \ 
$<\tau_{\, \rm R} (0,0)> \; \equiv \; <\tau_{\, \rm Ret} (0,0)>$ \ 
does not depend on the barrier width $a$;  \ (ii) in correspondence 
with the barrier region betwen $0$ and approximately $0.6 \ a$,
the value of $<\tau_{\, \rm Ret}(0,x)>$ is almost constant; \ while \ (iii)
its value increases with $x$ only in the barrier region near $x = a$ \ 
(even if
it should be pointed out that our calculations near $x = a$  are not so good,
due to the very small values assumed by \
$\int_{-\infty}^{\infty} J_{-}(x,t) \drm t$ \ therein). \ \ Let us notice that 
point (i), also 
observed firstly for quasi-monochromatic particles,$^{[\IIrm-2]}$
is as well in accordance with the results obtained in ref.[II-1] for arbitrary 
wave packets. \ Moreover, also points (ii) and (iii) can be explained by 
interference phenomena inside the barrier: if, near $x=a$, the initial 
returning--wave packet is almost totally quenched by the initial
incoming--wave packet, then only a negligibly small piece of its
back edge (consisting of the components with the smallest velocities) does 
remain. \ With decreasing $x$
\ ($x \rightarrow 0$), \ the unquenched part of the returning--wave packet
seems to become more and more large (containing more and more rapid
components),
thus making the difference \ $<\tau_{\, \rm Ret}(0,x)> -
<\tau_{\, \rm Pen}(0,x)>$ \ almost constant. \ 
And the interference  between incoming and reflected waves at points 
$x \le 0$ does effectively constitute a retarding phenomenon \ [so that \  
$<t_-(x=0)>$ is larger than $<\tau_{\, \rm R}(x=0)$], \ which can explain 
the larger values of $<\tau_{\, \rm R}(x=0,x=0)>$ in comparison    with 
$<\tau_{\rm Tun}(x=0,x=a)>$.

\h Therefore our evaluations, in all the cases considered above,
appear to confirm our previous analysis  
at page 352 of ref.[II-1], and our conclusions therein concerning in particular 
the validity of the Hartman effect also for {\em non}--quasi-monochromatic 
wave packets. \ Even more, since the interference between incoming and
reflected waves before the barrier (or between penetrating and returning 
waves, inside the barrier, near the entrance wall) does just {\em increase} 
the tunnelling time as well as the transmission times,
we can expect that our non-relativistic formulae for  
$<\tau_{\rm Tun}(0,a)>$  and  $<\tau_{\rm T}(x_{\rm i} < 0, x_{\rm f} > a)>$
will always forward positive values.$^{\# 3}$\\
\footnotetext{$^{\# 3}$ A different claim by Delgado, Brouard and 
Muga$^{[\IIrm-17]}$ does not seem to be relevant to our calculations, since it is 
based once more, like ref.[II-8],  not on our but on different wave packets (and
over--barier components are also retained in ref.[II-17], at variance 
with us). \  Moreover, in their classical example, they overlook the fact that 
the mean entrance time $<t_+(0)>$ gets contribution  mainly by the rapid 
components of the wave packet; they forget, in fact, that the slow components 
are (almost) totally reflected by the
initial wall, causing a quantum--mechanical reshaping that contributes to the 
initial ``time decrease" discussed by us already in the last few paragraphs 
of page 352 in ref.[II-1]. \ All such phenomena {\em reduce} the value of 
$<t_+(0)>$, and we expect it to be (in our non-relativistic treatment) 
less than  $<t_+(a)>$.}

\h At this point, it is necessary ---however--- to observe the following. 
 \ {\em Even if our non-relativistic equations are not expected (as we have 
just seen) to yield negative times}, {\bf nevertheless
one ought to bear in mind that (whenever
it is met an object}, ${\cal O}$, {\bf travelling at Superluminal speed) 
negative contributions should be expected 
to the tunnelling times: and this ought not to be
regarded as unphysical}.  In fact, whenever an ``object" ${\cal O}$ {\em 
overcomes}
the infinite speed$^{[\IIrm-18]}$ with respect to a certain observer, 
it will afterwards appear to the same observer as its ``{\em anti}-object"
$\ove{\cal O}$ travelling in the opposite {\em space} direction$^{[\IIrm-18]}$.  \
For instance, when passing from the lab to a frame ${\cal F}$ moving in 
the {\em same}
direction as the particles or waves entering the barrier region, the 
objects $\cal O$ penetrating through the final part of the barrier (with 
almost infinite speeds, like in Figs.II-1--5) 
will appear in the frame ${\cal F}$ as anti-objects $\ove{\cal O}$ 
crossing that portion of the barrier {\em in the opposite 
space--direction}$^{[\IIrm-18]}$.  In the new frame ${\cal F}$,
therefore, such anti-objects $\ove {\cal O}$ would yield a 
{\em negative} contribution to the
tunnelling time: which could even result, in total, to be negative. \ 
For any clarifications, see refs.[II-18]. \ So, we have no
objections a priori against the fact that Leavens can find, in certain cases, 
negative values$^{[\IIrm-8,17]}$:
e.g, when applying our formulae to wave packets with suitable 
initial conditions. \ What we want to stress here is that the appearance
of negative$^{[\IIrm-2]}$ times (it being predicted by Relativity itself,$^{[\IIrm-18]}$  
when in presence
of {\em anything} travelling faster than $c$) is not a valid reason to rule 
out a theoretical approach.\\  

\h {\em At last}, let us ---incidentally--- recall and mention the following 
fact. Some preliminary 
calculations of penetration times (inside a rectangular barrier) 
for tunnelling gaussian wave packets had been presented by us in 1994 
in ref.[II-16].  Later on ---looking for
any possible explanations for the disagreement between the results in
ref.[II-8] and in our ref.[II-16]--- we discovered, 
however, that an exponential factor was missing in a term of one of the 
fundamental formulae on which
the numerical computations (performed by our group in Kiev) were based: a 
mistake that 
could not be detected, of course, by our careful checks about the computing 
process. \ \ Therefore, the new results of ours appearing here in Figs.II-1--2 
[and appeared in J. de Physique-I 5 (1995) 1351] should {\em replace} 
Figs.1--3 of ref.[II-16]. \ One may observe that, by using the same 
parameters as (or parameters very 
near to) the ones adopted by for the Figs.3 and 4 of ref.[II-8],
our new, corrected figures II-1 and 2 result to be more similar to Leavens' 
than the uncorrected ones (and this is of course a welcome step towards
the solution of the problem). \ One can verify once 
more, however, that  
our theory appears to yield {\em for those parameters} non-negative results
for $<\tau_{\, \rm Pen} (x_{\rm f})>$, contrarily to a claim in ref.[II-8]. 
 \ Actually, our previous general
conclusions have not been apparently affected by the mentioned mistake. \ In 
particular,
the value of $<\tau_{\, \rm Pen} (x_{\rm f})>$ increases with increasing 
$x_{\rm f}$, and tends to saturation for $x_{\rm f} \rightarrow a$. \
We acknowledge, however, that the difference in the adopted integration 
ranges  
($- \infty$ to $+ \infty$ for us, and $0$ to $+ \infty$ for ref.[II-8]) does
not play an important role, contrarily to our previous belief,$^{[\IIrm-16]}$ in
explaining the remaining discrepancy between our results and those in 
ref.[II-8]. Such a discrepancy {\em might} perhaps  depend on the fact
that the functions to be integrated do fluctuate heavily$^{\# 4}$ (anyway, 
\footnotetext{$^{\# 4}$  We can {\em only}
say that we succeeded in reproducing  results of the type put forth in 
ref.[II-8] by using larger steps; whilst the {\em ``non-causal"} results 
disappeared ---{\em in the considered cases}--- when adopting small enough 
integration steps.} 
we did carefully check that our own elementary integration step in the 
integration over $\drm k$ was small enough  in order to 
guarantee the stability of the numerical result, and, in particular, of
their sign, for strongly oscillating functions in the integrand). \ More
probably, the persisting disagreement can be merely due to the fact 
---as recently claimed also by Delgado et al.$^{[\IIrm-17]}$--- that 
{\em different} initial conditions for the wave packets 
were actually chosen in ref.[II-8] and in ref.[II-1].  \ Anyway,
our approach seems to get support, at least in some particular cases,
also by a recent article by Brouard et al., which ``generalizes" ---even if
starting from a totally different point of view--- some of our 
results.$^{[\IIrm-12,15]}$. 

\h Let us take advantage of the present opportunity for answering other
criticisms appeared in
ref.[II-8], where it has been furthermore commented about our way of
performing actual averages over the physical time.  We cannot agree with
those comments: let us re-emphasize in fact that, within conventional quantum
mechanics, the time $t(x)$ at which our particle (wave packet) passes through
the position $x$ is ``statistically distributed" with the probability
densities \ ${\rm d}t J_{\pm}(x,t) / \int_{-\infty}^{\infty} {\rm d}t
J_{\pm}(x,t)$, \ as we explained at page 350 of ref.[II-1]. \ This
distribution meets the requirements of the time--energy uncertainty
relation. 

\h We also answered in Sect.{\bf II-1} the comments in ref.[II-8] about
our analysis$^{[\IIrm-1]}$ of the dwell--time approaches.$^{[\IIrm-19]}$

\h The last object of the criticism in ref.[II-8] refers to the 
impossibility,
in our approach, of distinguishing between ``to be transmitted" and ``to be
reflected" wave packets at the leading edge of the barrier. \ Actually, we do
{\em distinguish} between them; only, we cannot ---of course--- 
{\em separate} them,
due to the obvious {\em superposition} (and interference) of both 
wave functions in
$\rho (x,t)$, in $J(x,t)$ and even in $J_{\pm} (x,t)$.  \ This is known to be
an unavoidable consequence of the superposition principle, valid for {\em wave
functions} in conventional quantum mechanics. \ That last objection, therefore,
should be addressed to quantum mechanics, rather then to us. \ \   
Nevertheless, Leavens' aim of comparing the definitions proposed
by us for the tunnelling times not only with conventional, but also with
non--standard quantum mechanics {\em might} be regarded a priori as 
stimulating and possibly worth of further investigation.

\vspace*{1 cm}

{ \bf II-4. -- Further remarks}

\vspace*{0.5 cm}

\h In connection with the question of ``causality" for relativistic 
tunnelling particles, let us stress that 
the Hartman-Fletcher phenomenon (very small tunnelling
durations), with the consequence of Superluminal velocities for 
sufficiently wide barriers, was found 
theoretically also in QFT for Klein--Gordon
and Dirac equations,$^{[\IIrm-1]}$ and experimentally for electromagnetic
evanescent--mode wave packets$^{[\IIrm-4-6]}$ (tunnelling photons). \ 
It should be recalled that the problem of
Superluminal velocities for electromagnetic wave packets in media
with anomalous dispersion, with absorption, or behaving as a barrier
for photons (such as regions with frustrated internal reflection)
has been present in the scientific literature since long (see, for 
instance, quotations [2,1,18], and refs. therein); even if 
a complete settlement of the causal problem ---already available for point
particles$^{[\IIrm-18]}$--- does not seem to be yet available for the case of
relativistic waves. \ 
Apparently, it is not sufficient to pay attention only to
group velocity and mean duration for a ``particle" passing through
a medium; on the contrary, it is important taking into 
account and studying ab initio the {\em variances} (and the higher order 
central moments) of the duration distributions, 
as well as the wave packet {\em reshaping} in presence of a barrier, or 
inside anomalous media (even if reshaping does {\em not} play {\em always} an 
essential role). 

\h Passing to the approaches {\em alternative} to the
direct description of tunnelling processes in terms of 
wave packets, let us here recall those ones 
which are based  on averaging over 
the set of all dynamical paths (through the Feynman path integral 
formulation, the
Wigner distribution method, and the non-conventional Bohm approach), and 
others
that use additional degrees of freedom  which can be used as ``clocks". 
General analyses of all such alternative approaches can be found in 
refs.[II-1,20--24] from different points of view. 

\h If one confines himself within the
framework of  conventional quantum mechanics, then the Feynman path
integral formulation seems to be adequate.$^{[\IIrm-24]}$ \ But it is not clear  
what procedure is
needed to calculate physical quantities within the Feynman--type 
approach$^{[\IIrm-23]}$,
and usually such calculations result in complex tunnelling durations.
The Feynman approach seems to need further modifications if one wants
to apply it to the time analysis of tunnelling processes, and its results
obtaind up to now cannot be considered as final.

\h As to the approaches based on introducing additional 
degrees of freedom
as ``clocks", one can often realize that the tunnelling time happens to be
noticeably distorted by the presence of such degrees of freedom. For example,
the B\"uttiker-Landauer time is connected with absorption or emission
of modulation quanta (caused by the time--dependent oscillating part 
of the barrier potential) during tunnelling, rather than with the 
tunnelling process itself.$^{[\IIrm-1,15]}$  \ And, with reference to the 
Larmor precession time, it has been
shown$^{[\IIrm-11,20]}$ that this time definition is connected not
only with the intrinsic tunnelling process, but also with the geometric 
boundaries
of the magnetic field introduced as a part of the clock: for instance, if 
the magnetic field region is 
infinite, one ends up with the phase tunnelling time, after an average 
over the (small) energy spread of the wave packet.  \ Actually, those
``clock" approaches, when applied to tunnelling wave packets, seem to lead
---after eliminating the distortion caused by the 
additional degrees  of freedom--- to the same results
as the direct wave packet approach, whatever be the weight function adopted
in the time integration.

\h A more pedagogical, and detailed, review on the same subject (in
Italian, however) has been ``published" electronically, as LANL Archive 
\# cond-mat/9802126.

\h Before closing this paper, in which we met Superluminal motions, we would
like to put forth and comment the following information.  Since the pioneering
work by Bateman, it is known that the relativistic wave equations
---scalar, electromagnetic and spinor--- admit solutions with subluminal 
($v < c$) group velocities[II-25]. \ More recently, also Superluminal ($v > c$)
solutions have been constructed for those homogeneous wave equations, in
refs.[II-26] and quite independently in refs.[II-27]:  in some cases
just by applying a Superluminal Lorentz transformation[II-18,28]. \ Exactly the
same happens in the case of acoustic waves, with the presence of ``sub-sonic" 
and  ``Super-sonic" solutions[II-29];  so that one can expect they to exist, e.g.,
also for seismic wave equations. Or, rather, we might expect the same to be true 
even in the case of gravitational waves. \ At last, it is interesting to remark
that some, at least, of the Super-sonic (and Super-luminal) solutions, 
when experimentally realized[II-30], appear to be X-shaped, so as predicted 
in 1982 in ref.[II-31].

\h A brief review on the experimental data, that ---in four different sectors
of physics--- seem to indicate the existence of Superluminal motions,
appeared as an Appendix to the paper in LANL Archives \# physics/9712051
(to be published in Physica A, 1998).\\

\

\
 
{\bf Acknowledgements:} The authors are deeply indebted, for their generous 
scientific help, to A.Agresti, G.Brown, V.L.Lyuboshitz, D.Stauffer, 
B.N.Zakhariev and A.K.Zaichenko (who performed also all the numerical 
evaluations). \ They thank also, for useful duscussions or cooperation, 
M.Baldo, S.Esposito, G.Giuffrida, H.Hern\'andez F., L.C.Kretly, 
V.L.Lyuboshitz, R.L.Monaco, 
J.G.Muga, G.Nimtz, T.V.Obikhod, E.C.Oliveira, E.Parigi, A.Ranfagni, 
G.Salesi, S.Sambataro, V.S.Sergeyev,  V.M.Shilov, M.T. Vasconselos, A.Vitale, 
Sir Denys Wilkinson, J.Vaz, and M.Zamboni--Rached.\\

\

\centerline{{\bf Captions of the Figures of Part II}}

\vspace*{1. cm}

{\bf Fig.II-1} -- Behaviour of the average penetration time 
$<\tau_{\, \rm Pen}(0,x)>$ 
(expressed in seconds) as a function of the penetration depth 
$x_{\rm f} \equiv x$ 
(expressed in {\aa}ngstroms) through a rectangular 
barrier with width $a = 5 \; {\rm \AA}$, for \ 
$\Delta k = 0.02 \; {\rm {\AA}}^{-1}$ (dashed line) 
 \ and \ $\Delta k = 0.01 \; {\rm {\AA}}^{-1}$ (continuous line), 
\ respectively. \ The other parameters are listed in footnote $\# 1$.
 \ It is worthwhile to notice that $<\tau_{\, \rm Pen}>$ rapidly increases for
the first, few initial {\aa}ngstroms ($\sim 2.5$ \AA), tending afterwards to a 
saturation value. This seems to confirm the existence of the so-called 
``Hartman effect".$^{[\IIrm-2,1,15]}$
 
\vspace*{0.6 cm}

{\bf Fig.II-2} -- The same plot as in Fig.II-1, for \ $\Delta k = 0.01 \; 
{\rm {\AA}}^{-1}$, \  except that now the barrier width
is $a = 10 \; {\rm \AA}$. \ Let us observe that the numerical values of the
(total) tunnelling time
$<\tau_{\rm T}>$ practically does not change when passing from \ 
$a = 5 \;$\AA  \ to \ $a = 10 \;$\AA, \ again in agreement with the
characteristic features$^{1}$
of the Hartman effect. \ Figures II-1 and 2 do improve (and correct) the 
corresponding ones, preliminarily presented by us in ref.[II-16].

\vspace*{0.6 cm}

{\bf Fig.II-3} -- Behaviour of $<\tau_{\, \rm Pen}(0,x)>$ (expressed in seconds) as a 
function of $x$ (expressed in {\aa}ngstroms), for tunnelling through a 
rectangular barrier with
width $a = 5 \;${\AA} and for different values of ${\ove E}$ and of 
$\Delta k$:

curve 1: $\Delta k = 0.02 {\rm \AA}^{-1}$ and $\ove E = 2.5 \; 
{\rm eV}$; \ \ 
curve 2: $\Delta k = 0.02 {\rm \AA}^{-1}$ and $\ove E = 5.0 \; 
{\rm eV}$; \ \ 
curve 3: $\Delta k = 0.02 {\rm \AA}^{-1}$ and $\ove E = 7.5 \; 
{\rm eV}$; \ \
curve 4: $\Delta k = 0.04 {\rm \AA}^{-1}$ and $\ove E = 5.0 \; 
{\rm eV}$.

\vspace*{0.6 cm}

{\bf Fig.II-4} -- Behaviour of $<\tau_{\, \rm Pen}(0,x)>$ (in seconds) as a 
function of $x$ (in {\aa}ngstroms)  for $\ove E = 5 \; {\rm eV}$ 
and different values of $a$ and $\Delta k$:

curve 1: $a = 5  \; {\rm \AA}$ and $\Delta k = 0.02 {\rm \AA}^{-1}$; \ \  
curve 2: $a = 5  \; {\rm \AA}$ and $\Delta k = 0.04 {\rm \AA}^{-1}$; \ \ 
curve 3: $a = 10 \; {\rm \AA}$ and $\Delta k = 0.02 {\rm \AA}^{-1}$; \ \ 
curve 4: $a = 10 \; {\rm \AA}$ and $\Delta k = 0.04 {\rm \AA}^{-1}$. 

\vspace*{0.6 cm}

{\bf Fig.II-5} --Behaviour of $<\tau_{\, \rm Ret}(x,x)>$ (in seconds) as a 
function of $x$ (in {\aa}ngstroms)  for 
different values of $a$,  $\ove E$ and $\Delta k$:

curve 1: $a = 5 \; {\rm \AA}$, \ $\ove E = 2.5 \; {\rm eV}$ and 
$\Delta k = 0.02 {\rm \AA}^{-1}$; \ \ 
curve 2: $a = 5 \; {\rm \AA}$, \ $\ove E = 5.0 \; {\rm eV}$ and 
$\Delta k = 0.02 {\rm \AA}^{-1}$; \ \
curve 3: $a = 5 \; {\rm \AA}$, \ $\ove E = 7.5 \; {\rm eV}$ and 
$\Delta k = 0.02 {\rm \AA}^{-1}$; \ \
curve 4: $a = 5 \; {\rm \AA}$, \ $\ove E = 2.5 \; {\rm eV}$ and 
$\Delta k = 0.02 {\rm \AA}^{-1}$; \ \
curve 5: $a = 5 \; {\rm \AA}$, \ $\ove E = 5.0 \; {\rm eV}$ and 
$\Delta k = 0.02 {\rm \AA}^{-1}$; \ \
curve 6: $a = 5 \; {\rm \AA}$, \ $\ove E = 7.5 \; {\rm eV}$ and 
$\Delta k = 0.02 {\rm \AA}^{-1}$; \ \
curve 7: $a = 5 \; {\rm \AA}$, \ $\ove E = 5.0 \; {\rm eV}$ and 
$\Delta k = 0.02 {\rm \AA}^{-1}$; \ \
curve 8: $a = 5 \; {\rm \AA}$, \ $\ove E = 5.0 \; {\rm eV}$ and 
$\Delta k = 0.02 {\rm \AA}^{-1}$.\\

\         

BIBLIOGRAPHY OF PART II:\\

{\bf References}\\

[II-1] V.S. Olkhovsky and E. Recami: Physics Reports {\bf 214} 
(1992) 339; which constitute the First Part of this review article (see
above).\hfill\break

[II-2] T.E. Hartman: J. Appl. Phys. {\bf 33} (1962) 3427; \ J.R. 
Fletcher: J. Phys. C{\bf 18} (1985) L55. \ See also C.G.B. Garret and D.E.
McCumber: Phys. Rev. A{\bf 1} (1970) 305; \ S. Chu and S. Wong: Phys. Rev. Lett.
{\bf 48} (1982) 738; \ S. Bosanac: Phys. Rev. A{\bf 28} (1983) 577; \ 
F.E. Low and P.F. Mende: Ann. of Phys. {\bf 210} (1991) 380.\hfill\break

[II-3] See, e.g., Th. Martin and R. Landauer: Phys. Rev. A{\bf 45}
(1992) 2611; \ R.Y. Chiao, P.G. Kwiat and A.M. Steinberg: Physica B{\bf 175}
(1991) 257; \ A. Ranfagni, D. Mugnai, P. Fabeni and 
G.P. Pazzi: Appl. Phys. Lett. {\bf 58} (1991) 774.\hfill\break

[II-4] A. Enders and G. Nimtz: J. Physique I {\bf 2} (1992) 1693; \ 
{\bf 3} (1993) 1089; \ Phys. Rev. B{\bf 47} (1993) 9605; 
 \ E{\bf 48} (1993) 632; \ J. Physique I {\bf 4} (1994) 1817; \
G. Nimtz, A. Enders and H. Spieker: J. Physique I {\bf 4}
(1994) 1; \ ``Photonic tunnelling experiments: Superluminal tunnelling", in
{\em Wave and particle in light and matter (Proceedings of the Trani 
Workshop, Italy, Sept.$\!$ 1992)}, ed. by A. van der Merwe and A. Garuccio
(Plenum; New York, in press); \ W. Heitmann and G. Nimtz: Phys. Lett. 
A{\bf 196} (1994) 154.\hfill\break

[II-5] A.M. Steinberg, P.G. Kwiat and R.Y. Chiao: Phys. Rev. Lett. 
{\bf 71} (1993) 708, and refs. therein; \ Scientific American {\bf 269} (1993) 
issue no.2, p.38. \ See also P.G. Kwiat, A.M. Steinberg, R.Y.Chiao, P.H. 
Eberhard and M.D. Petroff: Phys. Rev. A{\bf 48} (1993) R867; \ E.L. Bolda,
R.Y. Chiao and J.C. Garrison: Phys. Rev. A{\bf 48} (1993) 3890.\hfill\break

[II-6] A. Ranfagni, P. Fabeni, G.P. Pazzi and D. Mugnani: Phys. Rev. 
E{\bf 48} (1993) 1453; \ Ch. Spielmann, R. Szip\"ocs, A. Stingl and F. 
Krausz: Phys. Rev. Lett. {\bf 73} (1994) 2308. \ Cf. also J. Brown: New
Scientist (April, 1995), p.26.\hfill\break

[II-7] W. Jaworski and D.M. Wardlaw: Phys. Rev. A{\bf 37} (1988) 2843.
\hfill\break

[II-8] C.R. Leavens: Solid State Commun. {\bf 85} (1993) 115.\hfill\break

[II-9] R. Landauer and Th. Martin: Solid State Commun. {\bf 84} 
(1992) 115.\hfill\break

[II-10] R.S. Dumont and T.L. Marchioro: Phys. Rev. A{\bf 47} (1993) 85.
\hfill\break

[II-11] See {\em e.g.} V.S. Olkhovsky: Nukleonika {\bf 35} (1990) 99, and
refs. therein; in particular, V.S. Olkhovsky: Doctorate (Habilitation) Thesis,
Institute for Nuclear Research, Ukrainian Academy of Sciences, Kiev (1986). \
See also V.S. Olkhovsky, V.M. Shilov and B.N. Zakhariev: Oper. Theory Adv.
Appl. {\bf 46} (1990) 159.\hfill\break

[II-12] S.Brouard, R. Sala and J.G. Muga: Phys.Rev. A{\bf 49} (1994)
4312; \  some criticism to this paper appeared in  C.R. Leavens: Phys.
Lett. A{\bf 197} (1995) 88. \ Cf. also A.F.M. Anwar and M.M. Jahan: IEEE J.
Quantum Electronics {\bf 31} (1995) 3.\hfill\break

[II-13] L. Landau and E.M. Lifshitz: {\em Quantum Mechanics}, 
3rd ed. (Pergamon Press; Oxford, 1977), Sect.{\bf 20}.\hfill\break

[II-14] C.R. Leavens: Phys. Lett. A{\bf 197} (1995) 88.\hfill\break 

[II-15] V.S. Olkhovsky, E. Recami and A.K. Zaichenko: Report
INFN/FM--94/01 (Frascati, 1994). \ See also F. Raciti and G. Salesi:
J. de Phys. I {\bf 4} (1994) 1783.\hfill\break

[II-16]  V.S. Olkhovsky, E.Recami and A.K. Zaichenko: 
Solid State Commun. {\bf 89} (1994) 31.\hfill\break

[II-17] V. Delgado, S. Brouard and J.G. Muga: ``Does positive flux 
provide a valid definition of tunnelling time?", to appear in
Solid State Commun.\hfill\break 

[II-18] E. Recami: ``Classical tachyons and possible applications", 
Rivista Nuovo Cim. {\bf 9} (1986), issue no.6, pp.1-178; \ E. Recami:
``A systematic, thorough analysis of the tachyon causal paradoxes",
Found. of Phys. {\bf 17} (1987)  239-296; \ E.Recami: ``The Tolman--Regge 
{\em antitelephone} paradox: Its solution by tachyon dynamics", Lett. Nuovo
Cim. {\bf 44} (1985) 587.\hfill\break

[II-19] F. T. Smith: Phys. Rev. {\bf 118} (1960) 349; \
M. Buttiker: Phys. Rev. B{\bf 27} (1983) 6178; \ 
M.L. Goldberger and K.M. Watson: {\em Collision Theory} (Wiley;
New York, 1964); \
J.M. Jauch and J.P. Marchand: Helv. Phys. Acta {\bf 40} 
(1967) 217; \
Ph.A. Martin: Acta Phys. Austr. Suppl. {\bf 23} (1981) 157.\hfill\break

[II-20]  Z.H. Huang, P.H. Cutler, T.E. Feuchtwang, E. Kazes, H.Q. Nguen
and T.E. Sullivan, J. Vac. Sci. Technol. A{\bf 8} (1990) 186.\hfill\break

[II-21] C.R. Leavens and G.C. Aers: in {\em Scanning Tunnelling
Microscopy and Related Methods}, edited by  R.J. Behm, N. Garcia and 
H. Rohrer (Kluwer; Dordrecht, 1990), p.59.\hfill\break

[II-22] A.P. Jauho: in {\em Hot Carriers in Semiconductor nanostructures,
Physics and Applications}, edited by J. Shah ( Academic Press; Boston, 1992),
p.121.\hfill\break

[II-23] R. Landauer and Th. Martin: Rev. Mod. Phys. {\bf 66} (1994) 217.
 \ Cf. also E.H. Hauge and J.A. Stvneng: Rev. Mod. Phys. {\bf 71} 
(1989) 917.\hfill\break

[II-24] W. Jaworski and D.M. Wardlaw: Phys. Rev. A{\bf 48} (1993) 3375.\hfill
\break

[II-25] Bateman H., 
Electrical and Optical Wave Motion (Cambridge Univ. Press;
Cambridge, 1915). \ See also Brittingham J.N., {\em J. Appl. Phys.} {\bf 54}
(1983) 1179; \ Ziolkhowski R.W., {\em J. Math. Phys.} {\bf 26} (1985) 861; \
Durnin J., {\em J. Opt. Soc.} {\bf 4} (1987) 651; \ \ Barut A.O. et al., {\em
Phys. Lett. A} {\bf 143} (1990) 349; \ {\em Found. Phys. Lett.} {\bf 3} (1990)
303; \ {\em Found. Phys.} {\bf 22} (1992) 1267.\hfill\break

[II-26] Barut A.O. et al., 
{\em Phys. Lett. A} {\bf 180} (1993) 5; \ {\bf 189}
(1994) 277; \ S.Esposito: {\em Phys. Lett}. A225 (1997) 203; \
W.A.Rodrigues Jr. and J.Vaz Jr., ``Subluminal and
superluminal solutions in vacuum of the Maxwell equations and the
massless Dirac equation," to appear in  {\em Adv. Appl.
Cliff. Alg.}\hfill\break 

[II-27] See, e.g., Ziolkowski R.W., Besieris I.M. 
and Shaaravi A.M., {\em J. Opt.
Soc. Am. A} {\bf 10} (1993) 75; \ Donnally R. and Ziolkhowski R.W., {\em Proc. 
R. Soc. London A} {\bf 440} (1993) 541.\hfill\break
 
[II-28] See the first one of refs.[II-18] 
and refs. therein. \ See also Recami
E. and Rodrigues W.A., ``A model theory for tachyons in two dimensions", in
Gravitational Radiation and Relativity, ed. by Weber J. and Karade T.M. (World
Scient.; Singapore, 1985), pp. 151-203.\hfill\break

[II-29] Lu J-y and Greenleaf J.F., {\em IEEE Trans. Ultrason. Ferroelectr.
Freq. Control} {\bf 37} (1990) 438; \ {\bf 39} (1992) 19.\hfill\break

[II-30] Ziolkowski R.W., Lewis D.K. and 
Cook B.D., {\em Phys. Rev. Lett.} {\bf 62}
(1989) 147; \ Lu J-y and Greenleaf J.F., {\em IEEE Trans. Ultrason. 
Ferroelectr. Freq. Control} {\bf 39} (1992) 441; \ ``Nondiffracting 
electromagnetic X waves", submitted for pub.\hfill\break

[II-31] Barut A.O., Maccarrone G.D. and 
Recami E., {\em Nuovo Cimento A} {\bf 71} (1982) 509.\hfill\break

\end{document}